\documentclass{article}
\usepackage{amssymb}
\usepackage{amsmath}
\usepackage{epsf}
\usepackage{epsfig}
\usepackage{floatflt}
\usepackage{float}

\input{tcilatex}
\begin{document}

\begin{center}

{\Huge Generations of \textit{solvable} \textit{discrete-time} dynamical
systems}

\bigskip

\textbf{Oksana Bihun}$^{\left( a,1\right) }$, \textbf{Francesco Calogero}$%
^{\left( b,c,2,3\right) }$\bigskip

$^{a}~$Department of Mathematics, University of Colorado, Colorado Springs,

1420 Austin Bluffs Pkway, Colorado Springs, CO 80918, USA

$^{1}~$obihun@UCCS.edu

$^{b}~$Physics Department, University of Rome \textquotedblleft La
Sapienza\textquotedblright , Italy

$^{c}~$Istituto Nazionale di Fisica Nucleare, Sezione di Roma, Italy

$^{2}~$francesco.calogero@roma1.infn.it, $^{3}~$%
francesco.calogero@uniroma1.it

\bigskip

\textit{Abstract}
\end{center}

A technique is introduced which allows to generate---starting from any 
\textit{solvable discrete-time} dynamical system involving $N$
time-dependent variables---new, generally \textit{nonlinear}, generations of 
\textit{discrete-time} dynamical systems, also involving $N$ time-dependent
variables and being as well \textit{solvable} by algebraic operations
(essentially by finding the $N$ zeros of explicitly known polynomials of
degree $N$). The dynamical systems constructed using this technique may also
feature large numbers of \textit{arbitrary} constants, and they need \textit{%
not} be autonomous. The \textit{solvable} character of these models allows
to identify special cases with remarkable time evolutions: for instance, 
\textit{isochronous} or\textit{\ asymptotically isochronous} \textit{%
discrete-time} dynamical systems\textit{.} The technique is illustrated by
a few examples. \bigskip


\section{Introduction}

The investigation of the evolution in \textit{discrete-time} of dynamical
systems---and in particular the identification of such models which are
amenable to exact treatments---has become an important area of mathematical
physics in the last one-two decades, see for instance the following review
papers and books: \cite{V1991,CN1999,S2003,R2006,BS2008}. The models under
consideration in this paper describe the evolution in the \textit{%
discrete-time} variable $\ell =0,1,2,...$ of an arbitrary number $N$ of
identical points moving in the \textit{complex} plane, the positions of
which are characterized by $N$ \textit{complex} coordinates, for instance $%
x_{n}\equiv x_{n}\left( \ell \right) $ or $y_{n}\left( \ell \right) $. Both
the equations of motion characterizing these models, and their solution,
only involve the \textit{algebraic} operation of finding the $N$ zeros of an
explicitly known $\ell $-dependent polynomial of degree $N$ in $z$. The
technique to identify such models is analogous to, but more flexible than,
the approach employed in \cite{C2011a,C2011b,CL2012,CL2013a,CL2013b,BCL2014}
to identify and discuss several \textit{solvable discrete-time} many-body
problems.

\textbf{Notation 1.1}. Above and hereafter (unless otherwise indicated)
indices such as $n,$ $m,$ $j$ run over the integers from $1$ to $N,$ with $N$
a given positive integer ($N\geq 2$). All quantities---except those
taking \textit{integer} values, such as the indices and the discrete-time $%
\ell $---are \textit{complex} numbers. Hereafter superimposed arrows denote $N$-vectors,
for instance the $N$-vector $\vec{y}\equiv \left(
y_{1},y_{2},...,y_{N}\right) $ features the $N$ components $y_{n}$, 
while the underlined notation $%
\underline{x}\equiv \left\{ x_{1},x_{2},...x_{N}\right\} $ denotes the 
\textit{unordered} set of $N$ (complex) numbers $x_{n}$. In the following we
will generally limit consideration to the \textit{generic} case in which the 
$N$ \textit{complex} numbers $y_{m}$ or $x_{n}$ are \textit{all different
among themselves}.

In the following it will be important to associate to an \textit{unordered}
set $\underline{x}$ the $N$-vectors the\ components of which correspond to a
specific ordering assignment of the $N$ elements $x_{n}$ of $\underline{x}.$
There are of course generally $N!$ different such vectors, and we will
hereafter use for them the notation $\vec{x}_{\left[ \mu \right] }\equiv
\left( x_{\left[ \mu \right] ,1},x_{\left[ \mu \right] ,2},...,x_{\left[ \mu %
\right] ,N}\right) ,$ with the index $\mu $ identifying a specific ordering
assignment of the $N$ numbers $x_{n}$, hence taking---as above---the $N$ 
\textit{integer} values from $1$ to $N!$ (since $N$ different objects can of
course be ordered in $N!$ different ways).

We also introduce the notation $\sum_{n_{1},n_{2},...,n_{m}=1}^{N}{}^{\ast }$
(note the appended star!) to denote a sum ranging from $1$ to $N$ over each
of the $m$ indices $n_{1},n_{2},...,n_{m}$, with the restriction that these $%
m$ indices be \textit{all different among themselves}.

Finally, we adopt the usual convention according to which \textit{a void sum
vanishes} and \textit{a void product equals unity}; note that this implies
that the sum $\sum_{n_{1},n_{2},...,n_{m}=1}^{N}{}^{\ast }$ \textit{vanishes
identically} if $m$ exceeds $N,$ $m>N$. $\blacksquare $

\textbf{Remark 1.1}. In the following we mainly focus on dynamical systems
characterized by \textit{first-order} equations of motion, say 
\begin{equation}
y_{m}\left( \ell +1\right) =f_{m}\left( \vec{y}\right) ~,
\end{equation}%
but occasionally below it will also be of interest to consider higher-order
equations of motion. $\blacksquare $

\textbf{Remark 1.2}. As in the previous models \cite%
{C2011a,C2011b,CL2012,CL2013a,CL2013b,BCL2014}, the \textit{solvable}
equations of motion identified below determine the $N$ points characterized
by the $N$ coordinates, say, $x_{n}\equiv x_{n}\left( \ell \right) $, as an 
\textit{unordered} set $\underline{{x}}\left( \ell \right) \equiv \left\{
x_{1}\left( \ell \right) ,x_{2}\left( \ell \right) ,...,x_{N}\left( \ell
\right) \right\} $, indeed generally as the $N$ zeros of an $\ell $%
-dependent polynomial $p_{N}\left( z;\ell \right) $ of degree $N$ in $z$;
hence these equations of motion are only deterministic inasmuch as they
identify (uniquely) the \textit{unordered} set $\underline{x}\left( \ell
+1\right) $ in terms of the \textit{unordered} set $\underline{{x}}\left(
\ell \right) $, but they do \textit{not} associate each coordinate $%
x_{n}\left( \ell \right) $ to a specific value of the index $n$ labeling it.
So these models describe $N$ \textit{indistinguishable} \textquotedblleft
particles\textquotedblright , the positions of which in the \textit{complex} 
$x$-plane at the discrete-time $\ell $ are characterized by the $N$
coordinates $x_{n}\left( \ell \right) $. A preferred association of these
coordinates to their labels might of course be provided at each step of the 
\textit{discrete-time} evolution by an argument of \textit{contiguity}, but
this is interesting only if the $N$ values $x_{n}\left( \ell +1\right) $ are
all adequately \textit{separated from each other}---in the complex $x$%
-plane---and each of them is adequately \textit{close to one and only one}
of the $N$ values $x_{m}\left( \ell \right) $ being themselves \textit{well} 
\textit{separated among each other}; more about this below. $\blacksquare $

The main idea of the previous models \cite{C2011a,C2011b,CL2012,CL2013a} was
to identify a solvable \textit{discrete-time} evolution of an $N\times N$
matrix and to then focus on the evolution of its $N$ eigenvalues. A more
straightforward approach---already employed in \cite{CL2013b} and \cite%
{BCL2014}---focussed directly on the \textit{solvable} evolution of a
polynomial $p_{N}\left( z;\ell \right) $ of degree $N$ in $z$ and of its $N$
zeros $x_{n}\left( \ell \right) $. The more convenient approach employed in
the present paper focusses more directly on the \textit{discrete-time}
evolutions of the $N$ \textit{coefficients} $y_{m}\left( \ell \right) $
respectively the $N$ \textit{zeros} $x_{n}\left( \ell \right) $ of a
time-dependent polynomial $p_{N}\left( z;\ell \right) $, by taking advantage
of a key formula  relating these evolutions, see below.
The \textit{solvable discrete-time} dynamical systems which can be
identified via this approach are described in the following Section 2 (with
a proof postponed to Appendix A in order to avoid interrupting the flow of
the presentation), and several examples are discussed in the subsequent
Section 3 and in Appendix B. In the last Section 4 we indicate to what
extent the findings reported in the present paper go beyond previously
reported results, and we tersely outline possible future developments.

\bigskip

\section{\textit{Solvable discrete-time} dynamical systems}
\label{sec2}

A main protagonist of our treatment is the time-dependent monic polynomial 
\begin{equation}
p_{N}\left( z;\ell \right) \equiv p_{N}\left( z;\vec{y}\left( \ell \right) ;%
\underline{x}\left( \ell \right) \right) =z^{N}+\sum_{m=1}^{N}\left[
y_{m}\left( \ell \right) ~z^{N-m}\right] =\dprod\limits_{n=1}^{N}\left[
z-x_{n}\left( \ell \right) \right] ~.  \label{Pol}
\end{equation}

\textbf{Remark 2.1}. The notation $p_{N}\left( z;\vec{y}\left( \ell \right) ;%
\underline{x}\left( \ell \right) \right) $ is of course \textit{redundant},
since to define this monic polynomial of degree $N$ in the variable\textbf{\ 
}$z$ at any time $\ell $ it is clearly sufficient to assign \textit{either}
its $N$ \textit{coefficients} $y_{m}\left( \ell \right) $ \textit{or} its $N$
\textit{zeros} $x_{n}\left( \ell \right) $: see (\ref{Pol}). Indeed, the $N$ 
\textit{coefficients} $y_{m}\left( \ell \right) $ are defined in terms of
the $N$ \textit{zeros }$x_{n}\left( \ell \right) $ by the standard formulas 
\begin{subequations}
\label{yx}
\begin{equation}
y_{m}\left( \ell \right) =\frac{\left( -1\right) ^{m}}{m!}\sigma _{m}\left( 
\underline{x}\left( \ell \right) \right) ~,
\end{equation}%
where, above and hereafter (see \textbf{Notation 1.1}),%
\begin{equation}
\sigma _{m}\left( \underline{x}\right)
=\sum\limits_{n_{1},n_{2},...,n_{m}=1}^{N}{}^{\ast}\ \left(
x_{n_{1}}~x_{n_{2}}~\cdot \cdot \cdot ~x_{n_{m}}\right) ~;  \label{sigma}
\end{equation}%
and conversely the $N$ \textit{zeros} $x_{n}\left( \ell \right) $ are
uniquely determined (but only \textit{up to permutations}) by the $N$ 
\textit{coefficients} $y_{m}\left( \ell \right) $, although of course 
\textit{explicit} formulas to this effect are only available for $N\leq 4$. $%
\blacksquare $

The main tool of our approach is the following key formula, implied by (\ref%
{Pol}) (for a proof see Appendix A): 
\end{subequations}
\begin{subequations}
\label{key}
\begin{equation}
\tprod\limits_{j=1}^{N}\left[ x_{n}\left( \ell +p\right) -x_{j}\left( \ell
\right) \right] +\sum_{m=1}^{N}\left\{ \left[ y_{m}\left( \ell +p\right)
-y_{m}\left( \ell \right) \right] ~\left[ x_{n}\left( \ell +p\right) \right]
^{N-m}\right\} =0~,  \label{keya}
\end{equation}%
which holds for every positive integer $p$. Note that this formula entails
that the $N$ values of the variables $x_{n}\left( \ell +p\right) $---for the 
$N$ values of the index $n$ in the range from $1$ to $N$---are the $N$ zeros
of the following polynomial of degree $N$ in the \textit{complex} variable $%
z $: 
\begin{equation}
\hat{p}_{N}\left( z;\ell \right) =\tprod\limits_{j=1}^{N}\left[
z-x_{j}\left( \ell \right) \right] +\sum_{m=1}^{N}\left\{ \left[ y_{m}\left(
\ell +p\right) -y_{m}\left( \ell \right) \right] ~\left( z\right)
^{N-m}\right\} ~.  \label{keyb}
\end{equation}

The merit of this formula---in either one of its equivalent versions~(\ref%
{keya}) or (\ref{keyb})---is to relate the \textit{discrete-time} evolution
of the $N$ \textit{zeros} $x_{n}\left( \ell \right) $ to the \textit{%
discrete-time} evolution of the $N$ \textit{coefficients} $y_{m}\left( \ell
\right) $, allowing to identify directly the equations of motion of \textit{%
new solvable} dynamical systems from those of \textit{known solvable}
dynamical systems. Indeed let us assume that the $N$ quantities $y_{m}\left(
\ell \right) $ evolve in the \textit{discrete-time} variable $\ell $
according to the following dynamical system: 
\end{subequations}
\begin{equation}
y_{m}\left( \ell +p\right) =f_{m}\left( \vec{y}\left( \ell \right) ,\vec{y}%
\left( \ell +1\right) ,\ldots ,\vec{y}\left( \ell +p-1\right) ;\ell \right)
~,  \label{SolvDynSysty}
\end{equation}%
where the $N$ functions $f_{m}$ are conveniently assigned so that this
system is \textit{solvable}. For instance this \textit{solvable} dynamical
system might be any one of those treated in the papers \cite%
{C2011a,C2011b,CL2012,CL2013a,CL2013b,BCL2014}. Note that, at this stage, we
do not exclude that the functions $f_{m}$ might feature an explicit
dependence on the discrete-time variable $\ell $, implying thereby that the
corresponding \textit{solvable }dynamical system (\ref{SolvDynSysty}) is
\textit{not} autonomous; there indeed exist \textit{nonautonomous} \textit{%
discrete-time} dynamical systems which are nevertheless \textit{solvable},
for instance the nonlinear system treated in \cite{CL2013b} or the rather
trivial \textit{decoupled linear} system (with $p=1$) 
\begin{subequations}
\begin{equation}
y_{m}\left( \ell +1\right) =g_{m}\left( \ell \right) ~y_{m}\left( \ell
\right) +h_{m}\left( \ell \right)  \label{ygh}
\end{equation}%
with $g_{m}\left( \ell \right) $ and $h_{m}\left( \ell \right) $ arbitrarily
assigned functions, the solution of which reads of course as follows (see 
\textbf{Notation 1.1}):%
\begin{equation}
y_{m}\left( \ell \right) =y_{m}\left( 0\right) ~\tprod\limits_{\ell ^{\prime
}=0}^{\ell -1}\left[ g_{m}\left( \ell ^{\prime }\right) \right] +\sum_{\ell
^{\prime \prime }=0}^{\ell -1}\left\{ \left[ \tprod\limits_{\ell ^{\prime
}=\ell ^{\prime \prime }+1}^{\ell -1}g_{m}\left( \ell ^{\prime }\right) %
\right] ~h_{m}\left( \ell ^{\prime \prime }\right) \right\} ~.
\end{equation}

It is then plain that the dynamical system characterized by the equations of
motion implied by (\ref{key}) and (\ref{SolvDynSysty}), the equations of
motion of which read as follows, 
\end{subequations}
\begin{subequations}
\label{NewSolSyst}
\begin{eqnarray}
\!\!\!\!\!\!\!\!\!\!\!\!\!\!\!\!\!\!\!\!\!\!\!\!&&\tprod\limits_{j=1}^{N}\left[ x_{n}\left( \ell +p\right) -x_{j}\left( \ell
\right) \right]   \notag \\
\!\!\!\!\!\!\!\!\!\!\!\!\!\!\!\!\!\!&&+\sum_{m=1}^{N}\left\{ \left[ f_{m}\left( \vec{y}\left( \ell \right) ,\ldots
,\vec{y}\left( \ell +p-1\right) ;\ell \right) -y_{m}\left( \ell \right) %
\right] ~\left[ x_{n}\left( \ell +p\right) \right] ^{N-m}\right\} =0  \label{7a}
\end{eqnarray}%
is as well \textit{solvable}. Note that (\ref{7a}) is equivalent to the
prescription that the $N$ updated values $x_{n}\left( \ell +p\right) $ of
the $N$ coordinates $x_{n}\left( \ell \right) $ coincide with the $N$ zeros
of the following polynomial equation of degree $N$ in $z$,%
\begin{equation}
\tprod\limits_{j=1}^{N}\left[ z-x_{j}\left( \ell \right) \right]
+\sum_{m=1}^{N}\left\{ \left[ f_{m}\left( \vec{y}\left( \ell \right) ,\ldots
,\vec{y}\left( \ell +p-1\right) ;\ell \right) -y_{m}\left( \ell \right) %
\right] ~z^{N-m}\right\} =0~.  \label{7b}
\end{equation}%
Of course in both of the last two formulas the $N$ components $y_{m}\left(
k\right) $ of the $N$-vector $\vec{y}\left( k\right) $, where $k=\ell ,\ell
+1,\ldots ,\ell +p-1$, should be replaced by their expressions, see (\ref{yx}%
), in terms of the $N$ components of the unordered set $\underline{x}\left(
k\right) $, namely  
\begin{equation}
y_{m}\left( k\right) =\frac{\left( -1\right) ^{m}}{m!}\sigma_m\left( \underline{x}(k)\right) ~.  \label{ymxn}
\end{equation}

The claim that this dynamical system (\ref{NewSolSyst}) is \textit{solvable}
is of course validated by the fact that the coordinates $x_{n}\left( \ell
\right) $, yielding its solution at time $\ell $, are the $N$ \textit{zeros}
of the polynomial (\ref{Pol}), of degree $N$ in $z$---hence they are
obtainable by an \textit{algebraic} operation---while the \textit{%
coefficients} $y_{m}\left( \ell \right) $ of this polynomial (\ref{Pol}) are
themselves obtainable by \textit{algebraic} operations, since the dynamical
system (\ref{SolvDynSysty}) characterizing the time-evolution of these
quantities is by assumption itself \textit{solvable}. In particular the 
\textit{initial-value} problem for this dynamical system can be
solved---for an arbitrary assignment of the initial data $x_{n}\left(
0\right) $---via the following 3 steps: (i) compute the corresponding $N$ 
\textit{initial} data $y_{m}\left( k\right) $ with $k=0,1,...,p-1$ of the
dynamical system (\ref{SolvDynSysty}) via (\ref{ymxn}); (ii) obtain the $N$
values $y_{m}\left( \ell \right) $ for $\ell \geq p$ via the \textit{solvable%
} dynamical system (\ref{SolvDynSysty}) with the initial data
computed in step (i); (iii) find the $N$ zeros $x_{n}\left( \ell \right) $
with $\ell \geq p$ of the polynomial $p_{N}\left( z;\ell \right) $, see (\ref%
{Pol}), defined by its $N$ \textit{coefficients} $y_{m}\left( \ell \right) $
as computed in step (ii).

Specific examples of \textit{solvable} dynamical systems (\ref{NewSolSyst})
will be exhibited and discussed in the following Section 3.

It is moreover plain that the usefulness of the key formula (\ref{key}) is
not limited to the identification of the new \textit{solvable} dynamical
system (\ref{NewSolSyst}) associated to the previously known \textit{solvable%
} dynamical system (\ref{SolvDynSysty}): it also opens the possibility to 
\textit{iterate}---albeit up to a limitation that we explain below---the
procedure we used above in order to obtain the \textit{new} solvable
dynamical system (\ref{NewSolSyst}) from the \textit{known} \textit{solvable}
system (\ref{SolvDynSysty}), by then using as known input in this approach
just the newly identified \textit{solvable} dynamical system (\ref%
{NewSolSyst}). And clearly this approach can be iterated over and over
again, yielding endless hierarchies of \textit{new} \textit{solvable}
dynamical systems, in analogy to what was recently done for \textit{continuous-time}
dynamical systems, see \cite{BC2015}.

But there is a significant difference with respect to the analogous
treatment in the continuous-time case, see \cite{BC2015} and \cite{BCal2016}.
The difference originates from the fact that in the \textit{continuous-time}
case the time evolution is in fact completely \textit{deterministic},
implying that in that context one is dealing with \textit{distinguishable}
particles: the assignment of the initial values $x_{n}\left( 0\right) $ of
the particle positions in the \textit{complex} $x$-plane at the initial time 
$t=0$ determines \textit{unambiguously}---thanks to the continuity of the
time-evolution of the particle coordinates $x_{n}\left( t\right) $ as
functions of the \textit{continuous-time} variable $t$---the particle
positions $x_{n}\left( t\right) $ for all future time $t$: say, the
coordinate $x_{1}\left( t\right) $ is the one that has evolved continuously
over time from the initial datum $x_{1}\left( 0\right) ,$ and likewise for
all values of the labels $n$ in the range from $1$ to $N$ identifying the
coordinates $x_{n}\left( t\right) $. Note that this is the case even though
the particle positions $x_{n}\left( t\right) $ are, also in that context,
identified with the \textit{zeros} of a time-dependent polynomial $%
p_{N}\left( z;t\right) $ of degree $N$ in the \textit{complex} variable $z$.

In the \textit{discrete-time} case treated in this paper one is instead
dealing---as explained above---with \textit{indistinguishable} particles.
This does not cause any problem for the definition and solution of the 
\textit{discrete-time} model~(\ref{NewSolSyst}) described just above, which characterizes the
evolution in the \textit{discrete-time} variable $\ell $ of the \textit{%
unordered} set of coordinates $\underline{x}\left( \ell \right) $. But of
course it raises an issue when we try to use that model to describe the
evolution of the \textit{coefficients} of a polynomial, since this set of numbers is
of course an \textit{ordered} set.

To explain what we mean in the simplest setting---and thereby also clarify
some relevant differences among the \textit{continuous-time} case treated in 
\cite{BC2015} and the \textit{discrete-time} case treated in the present
paper---let us now focus on equations of motion of \textit{first} order, i.
e. let us refer---in the remaining part of this Section 2---to the $p=1$
special case of the dynamical systems (\ref{SolvDynSysty}) and (\ref%
{NewSolSyst}) and of the key formula (\ref{key}). The interested reader will
have no difficulty to extend the following treatment to values of the \textit{%
integer} parameter $p$ larger than \textit{unity}.

In the following we will refer to the \textit{solvable} dynamical system (%
\ref{SolvDynSysty}) as the \textit{seed dynamical system} and to the 
\textit{new} dynamical system (\ref{NewSolSyst})---the \textit{solvable}
character of which has been detailed just above---as the \textit{generation
zero dynamical system}. We will also refer to polynomial~(\ref{Pol}) as the
\textit{generation zero} polynomial.

Let us then try and iterate the process used to obtain the \textit{%
generation zero dynamical system }(\ref{NewSolSyst}) from the \textit{seed system}
 (\ref{SolvDynSysty}). The idea ---following the analogous
treatment in the \textit{continuous-time} case \cite{BC2015}---is to
introduce a \textit{new} (monic, $\ell $-dependent) polynomial of degree $N$ in the 
\textit{complex} variable $z$ characterized by the property that its $N$
\textit{coefficients} evolve in the \textit{discrete time} $\ell $ as the solutions
of the \textit{solvable} \textit{generation zero system }(\ref{NewSolSyst}).
Specifically, we introduce the \textit{generation one polynomial} 
\end{subequations}
\begin{eqnarray}
&&p_{N}^{\left( \mu _{1}\right) }\left( z;\ell \right) \equiv p_{N}^{\left(
\mu _{1}\right) }\left( z;\vec{y}^{\left( \mu _{1}\right) }\left( \ell
\right) ;\underline{x}^{\left( \mu _{1}\right) }\left( \ell \right) \right)
=z^{N}+\sum_{m=1}^{N}\left[ y_{m}^{\left( \mu _{1}\right) }\left( \ell
\right) ~z^{N-m}\right]  \notag \\
&=&\dprod\limits_{n=1}^{N}\left[ z-x_{n}^{\left( \mu _{1}\right) }\left(
\ell \right) \right] ~,  \label{PolGen1}
\end{eqnarray}%
such that its coefficient vector $\vec{y}^{\left( \mu _{1}\right) }\left( \ell
\right) =\left( y^{(\mu_1)}_1(\ell), \ldots, y^{(\mu_1)}_N(\ell) \right)$ (see \textbf{Notation~1.1}) coincides with an appropriate ordering (prescribed by the index $\mu_1$) of
the \textit{unordered} solution set $\underline{x}\left( \ell \right) $ 
of the \textit{generation zero dynamical system}~(\ref{NewSolSyst}) (of course with $p=1$). Here the
index $\mu _{1}$ labels the possible orderings of the $N$ \textit{complex}
numbers $x_{1}(\ell ),x_{2}(\ell ),\ldots ,x_{N}(\ell ),$ implying the
definition of the $N$-vectors 
$$\vec{x}_{\left[ \mu _{1}\right] }\left( \ell
\right) \equiv \left( x_{\left[ \mu _{1}\right] ,1}\left( \ell \right) ,x_{%
\left[ \mu _{1}\right] ,2}\left( \ell \right) ...,x_{\left[ \mu _{1}\right]
,N}\left( \ell \right) \right) .$$ This index $\mu _{1}$ takes the $N!$
integer values from $1$ to $N!$ (see \textbf{Notation 1.1}), since there are 
$N!$ different permutations of $N$ different objects, implying of course the
existence of $N!$, generally different, $N$-vectors $\vec{x}_{\left[ \mu _{1}%
\right] }\left( \ell \right) $. And the prescription for $\vec{y}^{(\mu_1)}(\ell)$ mentioned just above,
characterizing the \textit{generation one polynomial} (\ref{PolGen1}), reads as follows:%
\begin{equation}
\vec{y}^{\left( \mu _{1}\right) }\left( \ell \right) =\vec{x}_{\left[ \mu
_{1}\right] }\left( \ell \right) ~,~~~y_{m}^{\left( \mu _{1}\right) }\left(
\ell \right) =x_{\left[ \mu _{1}\right] ,m}\left( \ell \right) ~.
\label{yxmu1}
\end{equation}

\textbf{Remark 2.2}. Let us reiterate that throughout our discussion we focus for
simplicity on the \textit{generic} case of monic polynomials of degree $N$
in the \textit{complex} variable $z$ the $N$ \textit{coefficients} of
which---and likewise the $N$ \textit{zeros} of which---are \textit{all
different among themselves}, and correspondingly on dynamical systems
describing the evolution in the \textit{discrete time} $\ell $ of $N$ points moving
in the \textit{complex} plane the positions of which at the same time $\ell $
never coincide. It is indeed plain that this is the \textit{generic}
situation, and it will be clear from the following treatment that the
occurrence of ``particle collisions''---the coincidence of two particle
positions at the same time---is an event that can only occur for a set of
initial data having \textit{vanishing measure} in the space of such data. 
$\blacksquare $

\textbf{Remark 2.3}. To allay any possible uneasiness about the notion that
a specific permutation labeled by an index $\mu $ in the range $1\leq \mu
\leq N!$ identifies a specific order of the $N$ elements $x_{n}$ of an 
\textit{unordered} set $\underline{x}$ of $N$ \textit{complex} numbers $x_{n}
$ let us provide an example of a procedure to do so. One begins by \textit{%
defining} a \textit{specific} ordering assignment of the \textit{unordered}
set $\underline{x}$ of $N$ \textit{different} \textit{complex} numbers $x_{n}
$, for instance that characterized by the (``increasing'') \textit{%
lexicographic} \textit{rule} stipulating that, of two \textit{complex}
numbers with \textit{different real} parts, the one with \textit{%
algebraically smaller real} part comes \textit{first}, and of two \textit{%
complex} numbers with \textit{equal real} parts, the one with \textit{%
algebraically smaller imaginary} part comes \textit{first}. After this
ordering of the \textit{unordered} set $\underline{x}$ is thus
established---defining the $N$-vector $\vec{x}_{[1]}$ with components $x_{%
\left[ 1\right] ,n}$---one can subsequently apply $N!$ standard sequential
reorderings---labeled by the index $\mu $ ranging from $2$ to $N!$%
---consisting in $N!$ permutations of the $N$ components $x_{[1],n}$ of the $%
N$-vector $\vec{x}_{[1]},$ these permutations being themselves labeled by
the value of the index $\mu $ according to some standard rule, for instance
the standard \textit{lexicographic} ordering of the $N!$ permutations of $N$
different objects (for $N=3$: $abc,acb,bac,bca,cab,cba$). Note that---see 
\textbf{Remark 2.2}---we always assume to deal with the \textit{generic}
case of (monic) polynomials of degree $N$ featuring $N$ \textit{different
zeros}, therefore defining new polynomials with $N$ \textit{different coefficients}
 as detailed above, see (\ref{yxmu1}). $%
\blacksquare $

The \textit{generation one dynamical systems} characterize the evolution in
the discrete time $\ell $ of the \textit{unordered} set $\underline{x}%
^{\left( \mu _{1}\right) }\left( \ell \right) \equiv \left\{ x_{1}^{\left(
\mu _{1}\right) }\left( \ell \right) ,x_{2}^{\left( \mu _{1}\right) }\left(
\ell \right) ,...,x_{N}^{\left( \mu _{1}\right) }\left( \ell \right)
\right\} $ of the $N$ zeros of the polynomial $p_{N}^{\left( \mu _{1}\right)
}\left( z;\ell \right) ,$ see (\ref{PolGen1}) with (\ref{yxmu1}). It is
plain---via the key formula (\ref{key}) (with $\ell =1$) and (\ref{yxmu1}%
)---that the equations of motion of these \textit{new} dynamical systems
read as follows: 
\begin{subequations}
\label{DynSystGen1}
\begin{eqnarray}
\!\!\!\!\!\!\!\!\!\!\!\!\!\!\!\!\!\!&&\tprod\limits_{j=1}^{N}\left[ x_{n}^{\left( \mu _{1}\right) }\left( \ell
+1\right) -x_{j}^{\left( \mu _{1}\right) }\left( \ell \right) \right]\notag\\
\!\!\!\!\!\!\!\!\!\!\!\!\!\!\!\!\!\!&&+\sum_{m=1}^{N}\left\{ \left[ x_{\left[ \mu _{1}\right] ,m}\left( \ell
+1\right) -x_{\left[ \mu _{1}\right] ,m}\left( \ell \right) \right] ~\left[
x_{n}^{\left( \mu _{1}\right) }\left( \ell +1\right) \right] ^{N-m}\right\}
=0~, \label{DynSystGen1a}
\end{eqnarray}%
i.e. they imply that the $N$ (updated) elements $x_{n}^{\left( \mu
_{1}\right) }\left( \ell +1\right) $ of the \textit{unordered} set $%
\underline{x}^{\left( \mu _{1}\right) }\left( \ell +1\right) \equiv \left\{
x_{1}^{\left( \mu _{1}\right) }\left( \ell +1\right) ,x_{2}^{\left( \mu
_{1}\right) }\left( \ell +1\right) ,...,x_{N}^{\left( \mu _{1}\right)
}\left( \ell +1\right) \right\} $ are the $N$ zeros of the following
polynomial equation of degree $N$ in the \textit{complex} variable $z$:%
\begin{equation}
\tprod\limits_{j=1}^{N}\left[ z-x_{j}^{\left( \mu _{1}\right) }\left( \ell
\right) \right] +\sum_{m=1}^{N}\left\{ \left[ x_{\left[ \mu _{1}\right]
,m}\left( \ell +1\right) -x_{\left[ \mu _{1}\right] ,m}\left( \ell \right) %
\right] ~z^{N-m}\right\} =0~.
\end{equation}%
Here of course the coordinates $x_{\left[ \mu _{1}\right] ,m}\left( \ell
\right) $ appearing in these equations (\ref{DynSystGen1}) are the solutions
of the \textit{generation zero dynamical system}, see (\ref{NewSolSyst}),
ordered according to the prescription characterized by the value of the
index $\mu _{1}$, as explained above (and see also below). Thus,
the quantities $x_{\left[ \mu _{1}\right] ,m}\left( \ell +1\right) $ are the 
$N$ zeros---ordered according to the prescription identified by the value of
the index $\mu _{1}$---of the polynomial 
\begin{equation}
\tprod\limits_{j=1}^{N}\left[ z-x_{j}\left( \ell \right) \right]
+\sum_{m=1}^{N}\left\{ \left[ f_{m}\left( \vec{y}\left( \ell \right) ;\ell
\right) -y_{m}\left( \ell \right) \right] ~z^{N-m}\right\} =0
\end{equation}%
(see (\ref{7b}) with $p=1$) where the $N$ quantities $y_{m}\left( \ell
\right) $ must of course be replaced by their expressions (\ref{ymxn}) in
terms of the $N$ coordinates $x_{n}\left( \ell \right) $ (the order of the
labeling being in this context irrelevant, see (\ref{ymxn})).

To interpret~(\ref{DynSystGen1}) as the \textit{evolution law} for the
\textit{unordered} set $\underline{x}^{(\mu_1)}(\ell)$,
we must recall that, via  (\ref{yxmu1}), $x_{[\mu_1],m}(\ell)=y_m^{(\mu_1)}(\ell)$ and, using relation~(\ref{yx}), make the following replacement in~(\ref{DynSystGen1}):
\begin{equation}
x_{[\mu_1],m}(\ell)=y_m^{(\mu_1)}(\ell)=\frac{(-1)^m}{m!} \sigma_m\left( \underline{x}^{(\mu_1)}(\ell)\right).
\label{Vietamu1}
\end{equation}

The new dynamical system~(\ref{DynSystGen1}) is \textit{solvable}, since its solution is provided by the $N$ \textit{%
zeros} of the \textit{generation one polynomial} $p_{N}^{\left( \mu _{1}\right) }\left( z;\ell
\right) $ given by (\ref{PolGen1})---i. e., by an algebraic operation. Of course, the generation one
 polynomial is itself obtainable by algebraic operations since its 
\textit{coefficients} $y_{m}^{\left( \mu _{1}\right) }=x_{\left[ \mu _{1}%
\right] ,m}\left( \ell \right) $ are given by the  permutation of the 
\textit{unordered} set $\underline{x}\left( \ell \right) $ associated with the index $\mu_1$, itself the
solution of the \textit{solvable} dynamical system (\ref{NewSolSyst}).

Because the evolution prescribed by the new dynamical system~(\ref{DynSystGen1}) depends on the assignment
of the ordering of the zeros of the \textit{generation zero polynomial}~(\ref{Pol}), see~(\ref{yxmu1}), we
 discuss several possibilities of making this assignment.

One possibility is to make a specific assignment for the ordering
prescription, once and for all, corresponding to a specific assignment of
the value of the index $\mu _{1}$: for instance one might assume that all
unordered sets $\underline{x}(\ell)$ be replaced by $N$-vectors $\vec{x}(\ell)$ the $N$
components of which are ordered, say, \textit{lexicographically}. This has
the following consequences: (i) consideration is then limited to only a
specific one out of the  \textit{a priori }possible \textit{generation
one dynamical systems}; (ii) one is then dealing with a dynamical system
describing the evolution of \textit{distinguishable} particles, the identity
of which is identified by their relative positions in the \textit{complex} $%
x $-plane; (iii) the initial data for this specific dynamical system cannot
be freely assigned: indeed, their values must be assigned not only
consistently with their identities (which is always possible by adjusting
the identities to the assigned values), they must moreover be the $N$ 
\textit{zeros} of a polynomial of degree $N$ in its \textit{complex}
variable $z$ the $N$ \textit{coefficients} of which satisfy themselves the
assigned prescription, and this of course entails a limitation on the
corresponding $N$ \textit{zeros}, hence on these \textit{initial values}. 
Note that an equivalent way to describe this possibility is to let the 
\textit{initial data} be assigned \textit{arbitrarily} and then to adjust
the ordering assignment of the \textit{unordered} sets of data consistently
with this initial assignment: but then the very dynamics associated with
this point of view would be dependent on the assignment of the \textit{%
initial data}, which is not consistent with what is usually meant by the
definition of a dynamical system as a set of rules---not themselves
dependent on the initial data---which determine how the initial data evolve
over time...

Another possibility would be to assign an $\ell $-dependent ordering
prescription based on \textit{contiguity over time}: given two sets of
unordered data $\underline{x}\left( \ell \right) $ and $\underline{x}\left(
\ell +1\right) $ one might require that they be ordered so that the distance
in the \textit{complex} $x$-plane between these coordinates at time $\ell $
and $\ell +1$ having the \textit{same} label be \textit{less} than the
distance between the coordinates having different labels, $\left\vert
x_{n}\left( \ell +1\right) -x_{n}\left( \ell \right) \right\vert <\left\vert
x_{n}\left( \ell +1\right) -x_{m}\left( \ell \right) \right\vert $ if $n\neq
m$. Clearly this prescription defines \textit{unambiguously }the label
assignments (i. e., the particle identities) at time $\ell +1$ corresponding
to any label assignment at time $\ell $, and viceversa, for any \textit{%
generic} configuration of $2N$ points $x_{n}\left( \ell \right) $ and $%
x_{n}\left( \ell +1\right) $ in the \textit{complex} $x$-plane (i. e., for
any \textit{arbitrary} configuration excluding a set of configurations
having \textit{zero measure} in the set of all possible configurations); but
it is plain that this is a reasonable prescription only if $\left\vert
x_{n}\left( \ell +1\right) -x_{n}\left( \ell \right) \right\vert
<<\left\vert x_{n}\left( \ell +1\right) -x_{m}\left( \ell \right)
\right\vert $ if $n\neq m$, namely when---with this assignment---the
positions of every particle at time $\ell $ and $\ell +1$ are much closer
to each other than the positions of any two different particles among themselves at time 
$\ell $ and at time $\ell +1$ (see examples below).

A third interesting possibility---not discussed any further in the present
paper---is to assign in a \textit{random} manner---at every step of the 
\textit{discrete-time} evolution---the prescription to go from the \textit{%
unordered} set of the $N$ \textit{zeros }of the \textit{generation zero}
polynomial to the \textit{ordered} set of the $N$ \textit{coefficients} of
the \textit{generation one} polynomial, thereby producing a dynamical system
featuring a \textit{random} evolution.

Up to now we have discussed the \textit{first} iteration of our
approach---and this justified our notation $\mu _{1}$ rather than just $\mu $
for the relevant parameter identifying the prescription characterizing the
transition from the $N$ \textit{zeros} of the \textit{ generation zero
polynomial} to the $N$ \textit{coefficients} of the \textit{generation one
polynomial}, yielding the identification of \textit{solvable generation one
dynamical systems}. It is plain how further iterations could be performed,
yielding \textit{new solvable discrete-time dynamical systems}: these
developments---which are not detailed in the present paper---are rather
obvious given the analogies with the treatment provided in \cite{BC2015} in
the \textit{continuous-time} context and the new features of the \textit{%
discrete-time} context discussed above.

Let us end this Section 2 with the following important remark, which is then
illustrated by some of the examples discussed in the following Section 3.

\textbf{Remark 2.4}. A \textit{solvable} dynamical system may well inherit
some properties from the \textit{seed solvable} dynamical system generating
it. For instance, if the \textit{seed solvable} dynamical
system is \textit{isochronous} respectively \textit{asymptotically
isochronous} with period $L$, then the \textit{generation $k$ solvable} systems are \textit{isochronous} respectively \textit{asymptotically isochronous}  with  period $L$ or its multiple (at most $(N!)^k L$), for all $k=0,1,\ldots$
To elaborate, suppose that  the solution $\vec{y}(\ell)$ of the \textit{seed system} has the \textit{isochronicity} property $\vec{y}%
\left( \ell +L\right) =\vec{y}\left( \ell \right) $, where $L$ is a \textit{fixed positive integer}. Then the solution $\underline{x}(\ell)$ of the
\textit{generation zero} system  has the same property: $
\underline{x}\left( \ell +L\right) =\underline{x}\left( \ell \right) $. The solution $\underline{x}^{(\mu_1)}(\ell)$
of the \textit{generation one} system is also \textit{isochronous} with  period $L_1$:
$\underline{x}^{(\mu_1)}\left( \ell +L_1\right) =\underline{x}^{(\mu_1)}\left( \ell \right) $, where $L_1=L$ if
the \textit{lexicographic} rule (or one of its $N!$ variants) was used to order the zeros $\underline{x}(\ell)$ of the
\textit{generation zero} system and $L_1=pL$ with the positive integer $p\leq N!$ if the \textit{contiguity} rule was \textit{applied
instead}. Similarly, if the solution $\vec{y}(\ell)$ of the \textit{seed system} has the \textit{asymptotic isochronicity} property
 $\vec{y}\left( \ell +L\right) -\vec{y}\left( \ell \right)
\rightarrow 0$ as $\ell \rightarrow \infty $, then the solution $\underline{x}(\ell)$ of the
\textit{generation zero} system  has the same property: $\underline{x}\left( \ell +L\right) -\underline{x}\left( \ell
\right) \rightarrow 0$ as $\ell \rightarrow \infty $, while the solution $\underline{x}^{(\mu_1)}(\ell)$
of the \textit{generation one} system is \textit{asymptotically isochronous} with period $L_1$ that is an integer multiple of $L$, at most $N! L$.
$ \blacksquare $

The validity of this \textbf{Remark 2.4} is justified by considerations sufficiently analogous to those made in the \textit{continuous-time} context, see~\cite{BC2015}, not to warrant their repetition here. These considerations are also illustrated by examples reported in the following Section~3, in particular see Remark~3.1 in that section.

\bigskip

\section{Examples}
\label{sec3}

In this section we illustrate our findings via the treatment of some
examples.

\textbf{Notation 3.1}. In this section we often omit to indicate explicitly
the $\ell $-dependence of the quantities under consideration, and we use a
superimposed tilde to denote a unit updating of the discrete-time variable $%
\ell $, hence, for instance, $\tilde{y}_{m}\equiv y_{m}\left( \ell +1\right) 
$ and $\tilde{\tilde{y}}_{m}\equiv y_{m}\left( \ell +2\right)$. $%
\blacksquare $

Plots of solutions of the dynamical systems considered in this Section 3
are given in Appendix B.

\textbf{Example 1}. As \textit{seed dynamical system} take the simple
first-order dynamical system 
\end{subequations}
\begin{subequations}
\begin{equation}
\tilde{y}_{m}=a_{m}~y_{m}+b_{m}~,
\end{equation}%
corresponding to (\ref{SolvDynSysty}) with $p=1$ and
\begin{equation}
f_{m}\left( \vec{y}\right) =a_{m}~y_{m}+b_{m}~.
\label{12b}
\end{equation}%
\label{SystSeed}
\end{subequations}
Note that this dynamical system is a simpler version of (\ref{ygh}) because
the $2N$ parameters $a_{m}$ and $b_{m}$ are assumed to be $\ell $%
-independent.

It is easily seen that the solution of these equations of motion reads 
\begin{equation}
y_{m}\left( \ell \right) =\left( a_{m}\right) ^{\ell }~y_{m}\left( 0\right)
+ \left[ \frac{\left( a_{m}\right) ^{\ell }-1}{a_{m}-1}\right]
~b_{m}~.  \label{seed}
\end{equation}

The \textit{generation zero} system  
constructed from the \textit{seed system}~(\ref{SystSeed}) via the method described in Section~\ref{sec2}, see~(\ref{7a}), reads
\begin{eqnarray}
\!\!\!\!\!\!\!\!\!\!\!\!\!\!\!\!\!\!\!\!\!\!\!\!&&\tprod\limits_{j=1}^{N}\left[ x_{n}\left( \ell +1\right) -x_{j}\left( \ell
\right) \right]   \notag \\
\!\!\!\!\!\!\!\!\!\!\!\!\!\!\!\!\!\!&&+\sum_{m=1}^{N}\left\{ \left[(a_m-1)y_{m}\left( \ell \right) +b_m
\right] ~\left[ x_{n}\left( \ell +p\right) \right] ^{N-m}\right\} =0 ,
\label{SystGen0N}
\end{eqnarray}
where $y_m(\ell)$ are expressed in terms of $\underline{x}(\ell)$ via~(\ref{yx}).
Of course, this system describes  the evolution of the \textit{unordered} set $\underline{x}(\ell)$ of the zeros of the monic polynomial with the coefficients ${y}_m(\ell)$ given by~\eqref{seed}, see the discussion below for the case where $N=2$.

\textbf{Remark 3.1}. It is plain that, if 
\begin{equation}
a_{m}=\exp \left( \frac{2~\pi ~\mathbf{i}~q_{m}}{p_{m}}\right) ~,  \label{am}
\end{equation}%
with $p_{m}$ an arbitrary \textit{positive integer}, $q_{m}$ an arbitrary 
\textit{integer }(of course, \textit{coprime} to $p_{m}$), and of course $%
\mathbf{i,}$ above and hereafter, the imaginary unit so that $\mathbf{i}%
^{2}=-1$, then $y_{m}\left( \ell \right) $ is \textit{isochronous} with
period $L_{m}=p_{m}$, namely, for any arbitrary initial datum $y_{m}\left(
0\right) $, there holds the periodicity property $y_{m}\left( \ell
+L_m\right) =y_{m}\left( \ell \right) ~$. Note that (\ref{am}) implies $%
\left\vert a_{m}\right\vert =1$. While if $\left\vert a_{m}\right\vert <1,$
then clearly---again, for any arbitrary initial datum $y_{m}\left( 0\right) $%
---there obtains the asymptotic limit $\lim\limits_{\ell \to \infty}y_{m}\left( \ell \right)
=b_{m}/\left( 1-a_{m}\right) $.

And it is
as well plain that the \textit{generic} solutions of the \textit{generation zero model}~(\ref{SystGen0N})
obtained from
this \textit{seed model} (\ref{SystSeed}) via the technique described in Section~\ref{sec2}
have the following remarkable properties: (i)~they are \textit{isochronous}
with period $L=P$, where $P$ is the \textit{Least Common Multiple} of the
positive integers $p_{1},\ldots ,p_{N}$, provided that all the parameters $%
a_{1},\ldots ,a_{N}$ satisfy condition (\ref{am}); (ii)~they are \textit{%
asymptotically isochronous} with  asymptotic period $L=P$ if some (at
least one but not all) of the parameters $a_{1},\ldots ,a_{N}$, say, $a_{j}$%
, have modulus less than \textit{unity}, $|a_{j}|<1$, and the remaining
parameters, say, $a_{m_{1}},\ldots ,a_{m_{k}}$ satisfy  condition (\ref%
{am}); in this case, $P$ is the the \textit{Least Common Multiple} of the
indices $p_{m_{1}},\ldots ,p_{m_{k}}$; (iii)~they converge asymptotically to
fixed values (independent of the initial data) if $\left\vert
a_{m}\right\vert <1$ for all $m=1,\ldots ,N$; (iv)~if some among the
parameters $a_{1},\ldots ,a_{N}$, say, $a_{j}$, have modulus larger than
unity, $|a_{j}|>1$, for \textit{generic} initial data some of the $N$ coordinates $%
x_{n}(\ell )$ \textit{diverge asymptotically} as $\ell \rightarrow \infty $ and
others converge to a \textit{finite value}, as implied by the findings reported in
Appendix G (\textquotedblleft Asymptotic behavior of the zeros of a
polynomial whose coefficients diverge exponentially\textquotedblright ) of 
\cite{C2001}. $\blacksquare $

Let us discuss the \textit{generation zero} system that stems from the \textit{seed system}~(\ref{SystSeed}), for the case where $N=2$. To construct this \textit{generation zero} system, consider
the \textit{generation zero} polynomial given by 
\begin{equation}
p_{2}(z)=z^{2}+y_{1}(\ell )z+y_{2}(\ell )=\left[ z-x_{1}(\ell )\right]
\left[ z-x_{2}(\ell )\right] ,
\end{equation}%
where $y_{1}(\ell ),y_{2}(\ell )$ are given by~(\ref{seed}) with $N=2$. Note
that the \textit{ordered pair} of the coefficients $\vec{y}(\ell )=\left(
y_{1}(\ell ),y_{2}(\ell )\right) $ determines the \textit{unordered set }of
the zeros $\underline{x}(\ell )=\left\{ x_{1}(\ell ),x_{2}(\ell )\right\} $.
The evolution of the latter \textit{unordered} set is given by the dynamical
system 
\begin{subequations}
\begin{equation}
\underline{x}(\ell +1)=\left\{ g_{1}\left( \underline{x}(\ell )\right)
,g_{2}\left( \underline{x}(\ell )\right) \right\} ,
\end{equation}%
where 
\begin{eqnarray}
&&g_{m}(\underline{x})\equiv g_{m}\left( \{x_{1},x_{2}\}\right) =\frac{1}{2}%
\left[ a_{1}~\left( x_{1}+x_{2}\right) -b_{1}\right]  \notag \\
&&+\frac{(-1)^{m}}{2}\left\{ \left[ a_{1}~\left( x_{1}+x_{2}\right) -b_{1}%
\right] ^{2}-4~b_{2}-4~a_{2}~x_{1}~x_{2}\right\} ^{1/2}~.
\end{eqnarray}%
\label{SystGen0}
\end{subequations}
Of course, this system~(\ref{SystGen0}) is the particular case of system~(\ref{SystGen0N}) for $N=2$.
Its solution  with the initial condition 
$\underline{x}(0)=\{x_{1}(0),x_{2}(0)\}$ reads
\begin{subequations}
\begin{equation}
\underline{x}(\ell )=\left\{ x_{1}(\ell ),x_{2}(\ell )\right\} ,
\end{equation}%
where 
\begin{equation}
x_{m }(\ell )=-\frac{1}{2}y_{1}(\ell )+(-1)^m \frac{1}{2}\left\{ \left[
y_{1}(\ell )\right] ^{2}-4y_{2}(\ell )\right\} ^{1/2}
\end{equation}%
and $y_{m}(\ell )$ are given by~(\ref{seed}) with $%
y_{1}(0)=-x_{1}(0)-x_{2}(0)$ and $y_{2}(0)=x_{1}(0)x_{2}(0)$. In more
expanded form, 
\begin{eqnarray}
&&x_{m }(\ell )=\frac{1}{2}\left\{ \left( a_{1}\right) ^{\ell }~\left[
x_{1}\left( 0\right) +x_{2}\left( 0\right) \right] -\left( \frac{\left(
a_{1}\right) ^{\ell }-1}{a_{1}-1}\right) ~b_{1}\right\}  \notag \\
&&+(-1)^m ~\frac{1}{2}\left\{ \left[ \left( a_{1}\right) ^{\ell }~\left[
x_{1}\left( 0\right) +x_{2}\left( 0\right) \right] -\left( \frac{\left(
a_{1}\right) ^{\ell }-1}{a_{1}-1}\right) ~b_{1}\right] ^{2}\right.  \notag \\
&&\left. -4~\left[ \left( a_{2}\right) ^{\ell }~x_{1}\left( 0\right)
~x_{2}\left( 0\right) +\left( \frac{\left( a_{2}\right) ^{\ell }-1}{a_{2}-1}%
\right) ~b_{2}\right] ~\right\} ^{1/2}~,  \notag \\
&&\ell =1,2,3,...~.  \label{xpm}
\end{eqnarray}
\label{SolGen0}
\end{subequations}

\textbf{Remark 3.2}. If the generic complex number $z$ is defined via its
modulus and phase as follows, $z=\left\vert z\right\vert ~\exp \left( 
\mathbf{i~}\varphi \right) $ with $0<\varphi \leq 2\pi $, its square-root $%
\sqrt{z}$ is of course defined up to a sign ambiguity, say $\sqrt{z}=\pm 
\sqrt{\left\vert z\right\vert }~\exp \left( \mathbf{i~}\varphi /2\right) $,
with $\sqrt{\left\vert z\right\vert }>0$. But, above and hereafter, we
assume for definiteness---irrelevant as this is---that the notation $z^{1/2}$
indicates a specific determination of the square-root of $z$, say that with 
\textit{positive real part,} and, \textit{if the real part vanishes, with
positive imaginary part}. $\blacksquare $

Most of the plots of the solutions  $\underline{x}(\ell )$   of~(\ref{SystGen0}) in Appendix B correspond to the 
\textit{lexicographic} ordering of the solution pair $\underline{x}(\ell )=\left\{x_{1}(\ell ),x_{2}(\ell )\right \}$, see \textbf{Examples 1a, 1b} there.  \textbf{Example 1c} uses instead the ordering by \textit{contiguity}.

\textbf{Example 2}. Let us now report the \textit{generation one  dynamical
systems} that stem from the seed system~(\ref{SystSeed}), again taking
$N=2$ for simplicity.  Recall that for $N=2$ the \textit{generation one} polynomials are given by 
\begin{equation}
p_{2}^{(\mu _{1})}(z)=z^{2}+y_{1}^{(\mu _{1})}(\ell )z+y_{2}^{(\mu
_{1})}(\ell )=\left[ z-x_{1}^{(\mu _{1})}(\ell )\right] ~\left[ 
z-x_{2}^{(\mu _{1})}(\ell ) \right] ~,~~~\mu _{1}=1,2,
\end{equation}%
see~(\ref{PolGen1}),
where the ordered pair $\vec{y}^{(\mu _{1})}(\ell )=\left( y_{1}^{(\mu
_{1})}(\ell ),y_{2}^{(\mu _{1})}(\ell )\right) $ equals an appropriately
ordered pair $x_{1}(\ell ),x_{2}(\ell )$. For example, we may choose the 
\textit{lexicographic} order of $x_{1}(\ell ),x_{2}(\ell )$ if $\mu _{1}=1$
and the other \textit{(anti-lexicographic}) order if $\mu _{1}=2$.

The dynamical system for the unordered pair $\underline{x}%
^{(\mu_1)}(\ell)=\left\{x^{(\mu_1)}_1(\ell), x^{(\mu_1)}_2(\ell) \right\}$
is then given by

\begin{subequations}
\begin{equation}
\underline{x}^{(\mu_1)}(\ell+1)=\left\{g_1^{(\mu_1)}\left(\underline{x}%
^{(\mu_1)}(\ell)\right), g_2^{(\mu_1)}\left(\underline{x}^{(\mu_1)}(\ell)%
\right)\right\},
\end{equation}
where 
\begin{eqnarray}
&&g^{(\mu_1)}_m(\underline{x})\equiv g_m\left(\{x_1, x_2\}\right)= -\frac{1}{%
2} \gamma_{-}^{(\mu_1)}\left(\underline{x} \right)  \notag \\
&&+\frac{1}{2} (-1)^m \left\{ \left[ \gamma_{-}^{(\mu_1)}\left(\underline{x}%
\right) \right]^2 -4 \gamma_{+}^{(\mu_1)}\left(\underline{x}\right)
\right\}^{1/2}  \notag \\
&&m=1,2~,~~~\mu _{1}=1,2~,
\end{eqnarray}
with the pair
\begin{equation}
\left( \gamma _{-}^{\left( \mu _{1}\right) }\left( \underline{x}\right),  \gamma _{+}^{\left( \mu _{1}\right) }\left( \underline{x}\right)\right) 
\end{equation}
being equal to the  ordering of the pair  
\begin{equation}
\left\{\alpha
\left( \underline{x}\right) - \left( -1\right) ^{\mu _{1}}~\beta \left( 
\underline{x}\right) ~, \alpha
\left( \underline{x}\right) + \left( -1\right) ^{\mu _{1}}~\beta \left( 
\underline{x}\right) \right\}
\end{equation}
that corresponds to the index $\mu_1$,
where
\begin{equation}
\alpha \left( \underline{x}\right) =\frac{1}{2}~\left[ a_{1}~\left(
-x_{1}-x_{2}+x_{1}~x_{2}\right) -b_{1}\right]~
\end{equation}%
and 
\begin{equation}
\beta \left( \underline{x}\right) = \left\{ \left[ \alpha \left( \underline{x%
}\right) \right] ^{2}-~b_{2}+ a_2 x_1 x_2(x_1+x_2)\right\} ^{1/2}~.
\end{equation}
\label{SystGen1}
\end{subequations}

The solution of dynamical system~(\ref{SystGen1}) with the initial condition 
$\underline{x}(0)=\{x_{1}(0),x_{2}(0)\}$ is given by 
\begin{subequations}
\begin{equation}
\underline{x}^{(\mu _{1})}(\ell )=\left\{ x_{1}^{(\mu _{1})}(\ell
),x_{2}^{(\mu _{1})}(\ell )\right\} ,
\end{equation}%
where 
\begin{equation*}
x_{m }^{(\mu _{1})}(\ell )=-\frac{1}{2}y_{1}^{(\mu _{1})}(\ell )+(-1)^m \frac{1%
}{2}\left\{ \left[ y_{1}^{(\mu _{1})}(\ell )\right] ^{2}-4y_{2}^{(\mu
_{1})}(\ell )\right\} ^{1/2}
\end{equation*}%
and $y_{m}^{(\mu _{1})}(\ell )$ are given by formulas~(\ref{xpm}) for $%
x_{m }(\ell )$, with $x_{m}(0)$ replaced by $y_{m}^{(\mu _{1})}(0)$, where 
\begin{eqnarray}
y_{1}^{(\mu _{1})}(0) &=&-x_{1}(0)-x_{2}(0),  \notag \\
y_{2}^{(\mu _{1})}(0) &=&x_{1}(0) ~x_{2}(0).  \label{ymu1initial}
\end{eqnarray}%
\label{SolGen1}
\end{subequations}
As before, $\mu _{1}=1$ indicates the \textit{lexicographic} order of the
pair $y_{1}^{(\mu _{1})}(\ell ),y_{2}^{(\mu _{1})}(\ell )$, while $\mu
_{1}=2 $ indicates the other (\textit{anti-lexicographic}) order.

In summary, given the initial condition~$\{x_{1}(0),x_{2}(0)\}$, we can
solve dynamical system~(\ref{SystGen1}) as follows. \textit{First}, we order the
initial condition to ensure that the pair $(x_{1}(0),x_{2}(0))$ is in the
\textit{lexicographic} order. \textit{Second}, we compute $y_{m}^{(\mu _{1})}(0)$ by formulas~%
(\ref{ymu1initial}) and assign $\mu _{1}=1$ if the pair $\left(
-x_{1}(0)-x_{2}(0),x_{1}(0) x_{2}(0)\right) $ turns out to be in the 
\textit{lexicographic} order and $\mu _{1}=2$ otherwise. \textit{Third}, we compute $%
y_{m}^{(\mu _{1})}(\ell )$ using formulas~(\ref{xpm}) for $x_{m }(\ell )$,
with $x_{m}(0)$ replaced by $y_{m}^{(\mu _{1})}(0)$, while ensuring that
each pair $\left( y_{1}^{(\mu _{1})}(\ell ),y_{2}^{(\mu _{1})}(\ell )\right) 
$ is ordered according to the value of $\mu _{1}$ chosen in the previous
step. \textit{Fourth}, we compute $\underline{x}^{(\mu _{1})}(\ell )$ using formulas~%
(\ref{SolGen1}). In the plots of \textbf{Examples 2a,~2b} in Appendix B the solutions $\left(x_1^{(\mu_1)}(\ell),
x_2^{(\mu_1)}(\ell) \right)$ are ordered \textit{lexicographically}.

\textbf{Example 3.} We do not display the equations of motion of the
subsequent model with $k=2,$ nor the formulas displaying its solutions,
since they are not very illuminating and the interested readers can easily
figure them out for themselves. We rather display a few representative plots
of the solutions of the system in the $k=2$ generations for several
particular values of the parameters $a_m, b_m$ in the \textit{seed system}~%
(\ref{SystSeed}) with $N=2$, see Appendix B.

\textbf{Example 4}. In this example we take the following \textit{solvable
second-order discrete-time} dynamical system as the \textit{seed system}:
\begin{equation}
y_{m}(\ell +2)=a_{m}(\ell )~\frac{y_{m}^{2}(\ell +1)}{y_{m}(\ell )}%
+b_{m}(\ell )~y_{m}(\ell +1),
\label{SystEx4}
\end{equation}%
where $a_{m}(\ell )$ and $b_{m}(\ell )$ are some functions of $\ell $. Via
the substitution 
\begin{equation}
u_{m}(\ell )=\frac{y_{m}(\ell +1)}{y_{m}(\ell )}
\end{equation}%
we find that the solution of system~(\ref{SystEx4}) with the initial
conditions $y_{m}(0),y_{m}(1)$ is given by 
\begin{subequations}
\label{SolEx4}
\begin{equation}
y_{m}(\ell )=y_{m}(0)~\prod_{j=0}^{\ell -1}u_{m}(j),  \label{SolEx4a}
\end{equation}%
where 
\begin{equation}
u_{m}(\ell )=\frac{y_m(1)}{y_m(0)}~\prod_{j=0}^{\ell -1}a_{m}(j)+\sum_{k=0}^{\ell
-1}\left\{ \left[ \prod_{j=k+1}^{\ell -1}a_{m}(j)\right] b_{m}(k)\right\}.
\label{SolEx4b}
\end{equation}%

The evolution of the zeros of the polynomial $z^{N}+\sum\limits_{m=1}^{N}y_{m}(\ell
)z^{N-m}$ is then described by the dynamical system 
\end{subequations}
\begin{eqnarray}
\prod_{j=1}^{N}\left[ x_{n}(\ell +2)-x_{j}(\ell )\right] &&  \notag \\
+\sum_{m=1}^{N}\left\{ \left[ a_{m}(\ell )~\frac{\left[ y_{m}(\ell +1)\right]
^{2}}{y_{m}(\ell )}+b_{m}(\ell )~y_{m}(\ell +1)-y_{m}(\ell )\right] ~\left[
x_{n}(\ell +2)\right] ^{N-m}\right\} =0~, &&  \label{SystEx4x}
\end{eqnarray}%
where $y_{m}(\ell )$ and $y_{m}(\ell +1)$ in the right-hand side must be
replaced with the appropriate expressions depending on $\underline{x}(\ell )$
and $\underline{x}(\ell +1)$, see~(\ref{yx}).

Let us consider the  \textit{autonomous} case of system~(\ref{SystEx4}) with 
\begin{eqnarray}
&&a_m(\ell)=a_{m}~,\notag\\
&&b_m(\ell)=b_m
\label{ambmaut}
\end{eqnarray}
and
\begin{eqnarray}
a_{m}=\exp \left( 2\pi \mathbf{i}q_{m}/p_{m}\right)~,
\label{amPeriodic}
\end{eqnarray}
where $%
q_{m},~p_{m}$ \textit{coprime integers} and $p_{m}>0$.
In this case  system~(\ref{SystEx4}) reads
\begin{equation}
y_{m}(\ell +2)=a_{m}~\frac{y_{m}(\ell +1)^{2}}{y_{m}(\ell )}%
+b_{m}~y_{m}(\ell +1).
\label{SystEx4Casea}
\end{equation}%
and its solution is given by~(\ref{SolEx4a}), where 
\begin{equation}
u_{m}(\ell )=(a_{m})^{\ell }~u_{m}(0)+\frac{(a_{m})^{\ell }-1}{a_{m}-1}%
~b_{m}~  \label{SolEx4b}
\end{equation}%
and $u_m(0)=y_m(1)/y_m(0)$.
    Because $a_m$ are given by~(\ref{amPeriodic}), all $u_{m}(\ell )$ defined by the last formula are $%
P $-periodic, where $P$ is the Least Common Multiple of the $N$ integers $%
p_{m} $. It is then easy to see from~(\ref{SolEx4a}) that 
\begin{subequations}
\begin{equation}
y_{m}(\ell +P)=\alpha _{m}~y_{m}(\ell )~,
\end{equation}%
where 
\begin{equation}
\alpha _{m}=\prod_{k=0}^{P-1}u_{m}(k)~.  \label{alpham1}
\end{equation}%
\end{subequations}
Suppose that the initial data $\underline{x}(0),~\underline{x}(1)$  for system~(\ref{SystEx4x}) 
are such that  $u_m(\ell)$ given by~\eqref{SolEx4b} with
$
u_{m}(0)=y_{m}\left( 1\right) /y_{m}(0)=\sigma _{m}(\underline{x}(1))/\sigma
_{m}(\underline{x}(0))$ (see~(\ref{yx})) satisfy the condition
\begin{equation}
\prod_{k=0}^{P-1}u_{m}(k)=\rho _{m}\exp \left( 2\pi 
\mathbf{i}r_{m}/s_{m}\right),
\label{Conduk}
\end{equation} 
where $r_{m},s_{m}$ are \textit{coprime
integers} with $s_{m}>0$ and $\rho _{m}$ are \textit{positive real} numbers such that 
$\rho _{m}\leq 1$ with at least one of them being equal to \textit{unity}, say $%
 \rho _{j} =1$. That is, suppose that $\underline{x}(0),~\underline{x}(1)$ are such that
 \begin{subequations}
\begin{equation}
\prod_{k=0}^{P-1}\left[ (a_{m})^{k}~\frac{\sigma _{m}(\underline{x}(1))}{\sigma
_{m}(\underline{x}(0))}+\frac{(a_{m})^{k}-1}{a_{m}-1}%
~b_{m}\right] =\beta _{m},  \label{PC1}
\end{equation}%
where
\begin{equation}
\beta_m=\rho _{m}\exp \left( 2\pi 
\mathbf{i}r_{m}/s_{m}\right).
\label{alpham}
\end{equation} 
\label{PeriodicityConditions}
\end{subequations}
Then the solutions $y_{m}(\ell )$ of system~(\ref{SystEx4Casea}), (\ref{amPeriodic}) with the
initial data $y_m(k)=(-1)^m \sigma_m(\underline{x}(k))/m!$, $k=0,1$, are \textit{periodic} or \textit{asymptotically periodic} with  period~$%
L $ that is the Least Common Multiple of the integers in the set $%
\{s_{m}P: \rho_{m}=1\}$ (of course, the \textit{periodic} case corresponds to the situation where
all $\rho _{m}=1$).

To summarize, we conclude that the \textit{generation zero} system~(\ref{SystEx4x}) in the autonomuous case
given by~(\ref{ambmaut}), (\ref{amPeriodic}) features a periodic or an asymptotically periodic solution with 
period $L$ provided that
the \textit{initial} data $\underline{x}(0),\underline{x}(1) $
 satisfy conditions~(\ref{PeriodicityConditions}), see \textbf{Remark 2.4}. Plots of a periodic 
solution of~(\ref{SystEx4Casea}) are given in \textbf{Example 4} of Appendix B.

\bigskip

\section{Looking backward and forward}

In this last section we tersely indicate the extent to which the findings
reported in the present paper go beyond previously reported results, and we
outline possible future developments.

A basic idea underlies the identification of many, perhaps most, \textit{%
solvable }dynamical systems characterizing the time-evolution of $N$ points
moving in the \textit{complex} plane (or, equivalently, in the \textit{real}
Cartesian plane). The idea is to identify/manufacture dynamical systems 
\textit{solvable} by algebraic operations---hence evolving \textit{%
nonchaotically}---by taking advantage of the \textit{nonlinear} but \textit{%
algebraic} relations among the $N$ \textit{zeros} and the $N$ \textit{%
coefficients} of a time-dependent polynomial $p_{N}\left( z;t\right) $ of
degree $N$ in the complex variable $z$. Its exploitation is quite old---see
for instance \cite{C1978}---yet only quite recently a very useful tool to
better take advantage of this approach has emerged \cite{C2015a}, leading to
the identification of several \textit{new} solvable dynamical systems \cite%
{BC2015,C2015a,BC2015a,BC2015c,C2016,C2016a,C2016b,C2016c,C2016d}. These
developments---which also led to the idea of \textit{generations of
polynomials} \cite{BC2015}---focussed up to now on evolutions in the \textit{%
continuous-time} variable $t$. The main novelty of the present paper is the
extension of this approach to evolutions in the \textit{discrete-time}
variable $\ell $.

To conclude this terse outline of previous developments, let us insert the
following

\textbf{Remark 4.1}. A simple way to ``generalize'' any dynamical system is
to perform a change of dependent and independent variables; but generally
the ``new'' models obtained in such a way from a known model are not
considered \textit{really new}. It might therefore be inferred that the
techniques described in all the papers referred to above, and in the present
paper, are not really yielding \textit{new }solvable dynamical systems,
since the main tool employed---the relation between the \textit{coefficients}
and the \textit{zeros} of a polynomial---may well be considered just a
change of dependent variables for the dynamical systems under consideration.
But this criticism conflicts with the observation that essentially \textit{%
all solvable} dynamical system can be reduced, by \textit{appropriate}
changes of variables, to \textit{trivial} evolutions. The rub is, of course,
to identify the \textit{appropriate} changes of variables. Hence the
emergence of the ``inverse'' approach: to start from certain changes of
variables---in particular, those relating the (time-dependent) \textit{%
coefficients} and the (time-dependent) \textit{zeros} of (time-dependent,
monic) polynomials---and to then try and identify the dynamical systems 
\textit{solvable} via this kind of transformation of dependent variables. To
those who consider such an ``inverse'' approach a kind of cheating, we can
only reply by begging them to ponder what is written in the Foreword (see,
in particular, page VII) of the book \cite{C2001} to justify this approach
(indeed, amply practiced both in that book and in most other publications on 
\textit{solvable/integrable} dynamical systems, including the present one). $%
\blacksquare $

Let us finally look forward and tersely list a possible future development.
An analogous approach to that discussed in this paper---but leading to 
\textit{solvable} dynamical systems in \textquotedblleft $q$-discrete
time\textquotedblright , that is, characterized by $q$-difference equations
of motion rather than difference equations of motion---obtains by taking as
point of departure, instead of the polynomial formula (\ref{Pol}), the
following definition: 
\begin{equation}
p_{N}\left( z;q;\ell \right) =z^{N}+\sum_{m=1}^{N}\left[ y_{m}\left( q^{\ell
}\right) ~z^{N-m}\right] =\prod\limits_{n=1}^{N}\left[ z-x_{n}\left( q^{\ell
}\right) \right] ~,
\end{equation}%
with $q$ an arbitrary parameter (of course $q\neq 1$). It is then easily
seen---again, by a quite analogous treatment to that provided in Appendix
A---that the key formula (\ref{keya})  is replaced by the relation 
\begin{equation}
\prod_{j=1}^{N}\left[ x_{n}\left( q^{\ell +p}\right) -x_{j}\left( q^{\ell
}\right) \right] +\sum_{m=1}^{N}\left\{ \left[ y_{m}\left( q^{\ell
+p}\right) -y_{m}\left( q^{\ell }\right) \right] ~\left[ x_{n}\left( q^{\ell
+p}\right) \right] ^{N-m}\right\} =0~,
\end{equation}%
where $p$ is a positive integer.
And a simple example of a \textit{first order} \textit{seed} system to generate other nonlinear \textit{%
solvable} dynamical systems in \textquotedblleft $q$-discrete
time\textquotedblright\ reads as follows: 
\begin{subequations}
\begin{equation}
y_{m}\left( q^{\ell +1}\right) =a_{m}~y_{m}\left( q^{\ell }\right) +b_{m}~,
\end{equation}%
since the explicit solution of its \textit{initial-value} problem clearly
reads%
\begin{equation}
y_{m}\left( q^{\ell }\right) =\left( a_{m}\right) ^{\ell }~y_{m}\left(
1\right) +\left[ \frac{\left( a_{m}\right) ^{\ell }-1}{a_{m}-1}\right]
~b_{m}~.
\end{equation}

\bigskip

\section{Acknowledgements}

The first author (O.~Bihun) would like to acknowledge with gratitude the hospitality of the Physics Department of the ``La Sapienza'' University of Rome during this author's multiple visits. This paper was essentially finalized during the  visit in Summer 2016.

\section{Appendix A: Proof of formula (\protect\ref{key})}

Our task in this Appendix A is to prove key formula (\ref{key}), which
is in fact a rather immediate consequence of definition (\ref{Pol}) of
the polynomial $p_{N}\left( z;\ell \right) $. Let $p$ be a \textit{positive
integer}. The first of equalities (\ref{Pol}) implies 
\end{subequations}
\begin{subequations}
\label{AA}
\begin{equation}
p_{N}\left( z;\ell +p\right) -p_{N}\left( z;\ell \right)
=\sum_{m=1}^{N}\left\{ \left[ y_{m}\left( \ell +p\right) -y_{m}\left( \ell
\right) \right] ~z^{N-m}\right\} ~,  \label{Aa}
\end{equation}%
and, via the second of equalities (\ref{Pol}), this formula reads%
\begin{equation}
\dprod\limits_{j=1}^{N}\left[ z-x_{j}\left( \ell +p\right) \right]
-\dprod\limits_{j=1}^{N}\left[ z-x_{j}\left( \ell \right) \right]
=\sum_{m=1}^{N}\left\{ \left[ y_{m}\left( \ell +p\right) -y_{m}\left( \ell
\right) \right] ~z^{N-m}\right\} ~.  \label{Ab}
\end{equation}%
It is now plain that, by setting in this formula $z=x_{n}\left( \ell
+p\right) ,$ one obtains (\ref{keya}). Q. E. D.

\textbf{Remark A.1}. It is also plain that, by setting $z=x_{n}\left( \ell
\right) $ in (\ref{Ab}), one obtains the alternative formula 
\end{subequations}
\begin{eqnarray}
\!\!\!\!\!\!\!\!\!\!\!\!\!\!\!\!\!\!&&\tprod\limits_{j=1}^{N}\left[ x_{n}\left( \ell \right) -x_{j}\left( \ell
+1\right) \right] =\sum_{m=1}^{N}\left\{ \left[ y_{m}\left( \ell +1\right)
-y_{m}\left( \ell \right) \right] ~\left[ x_{n}\left( \ell \right) \right]
^{N-m}\right\} ~.
\end{eqnarray}
$\blacksquare$

\bigskip

\section{Appendix B: Plots of solutions of systems treated in Section~%
\protect\ref{sec3}}

In this Appendix B we provide several plots of the solutions $\underline{x}%
\left( \ell \right) = \{x_{1}(\ell ),x_{2}(\ell )\}$ of the \textit{%
discrete-time} dynamical systems treated in Section 3 for $N=2$ and for various
assignments of the parameters and of the initial conditions. Because the
solution $\underline{x}\left( \ell \right) = \{x_{1}(\ell ),x_{2}(\ell
)\}$ is an \textit{unordered} set, for the purpose of plotting it, we  assume that the pair
$\left(x_{1}(\ell ),x_{2}(\ell )\right)$ is ordered \textit{lexicographically}, 
with the only exception of \textbf{Example 1c}, in which the solution set $\underline{x}(\ell)$
is ordered by contiguity,
see \textbf{Remark 2.3}.
In the following graphs, the (dashed or continuous; if any) line segments
joining points are only visual aids having no other significance for the purpose of our discussion.

\textbf{Example 1. \textit{Generation zero} system with \textit{seed}~(\ref{SystSeed})}

Choose $N=2$ in the seed system~(\ref{SystSeed}) to obtain, via the method described in Section~\ref{sec2}, system~(\ref{SystGen0}) with  solution~(\ref{SolGen0}). Let us recall that \textbf{Remark 3.1} predicts the
behavior of solutions (\ref{SolGen0}) depending on the values of the parameters $a_{1},a_{2}.$

\begin{figure}[tbp]
\begin{center}
\includegraphics[width=10cm]{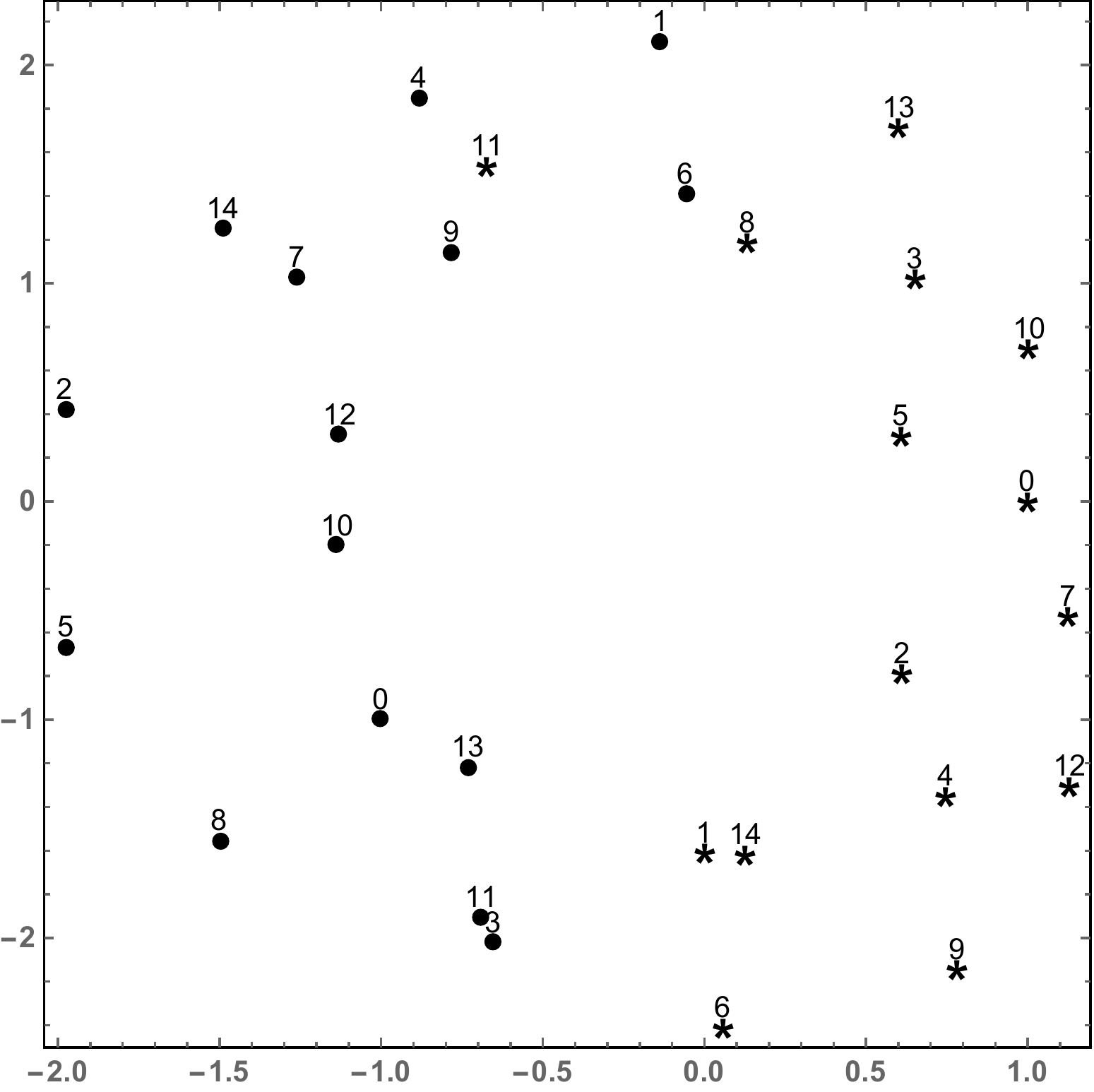}
\end{center}
\caption{\textbf{Example 1a.} For each $\ell=0,1,\ldots, 15$, the position of $%
x_{1}(\ell)$ in the \textit{complex} $x$-plane is indicated by a \textit{dot}
and the label $\ell$ that indicates the value of the discrete-time, while
the position of $x_{2}(\ell)$ is indicated by a \textit{star} and the
analogous label~$\ell$. The complex pairs~$\left(x_{1}(\ell), x_2(\ell)\right)$ are
ordered \textit{lexicographically}.}
\label{Fig:Ex1aBoth}
\end{figure}

\begin{figure}[tbp]
\begin{center}
\includegraphics[width=8cm]{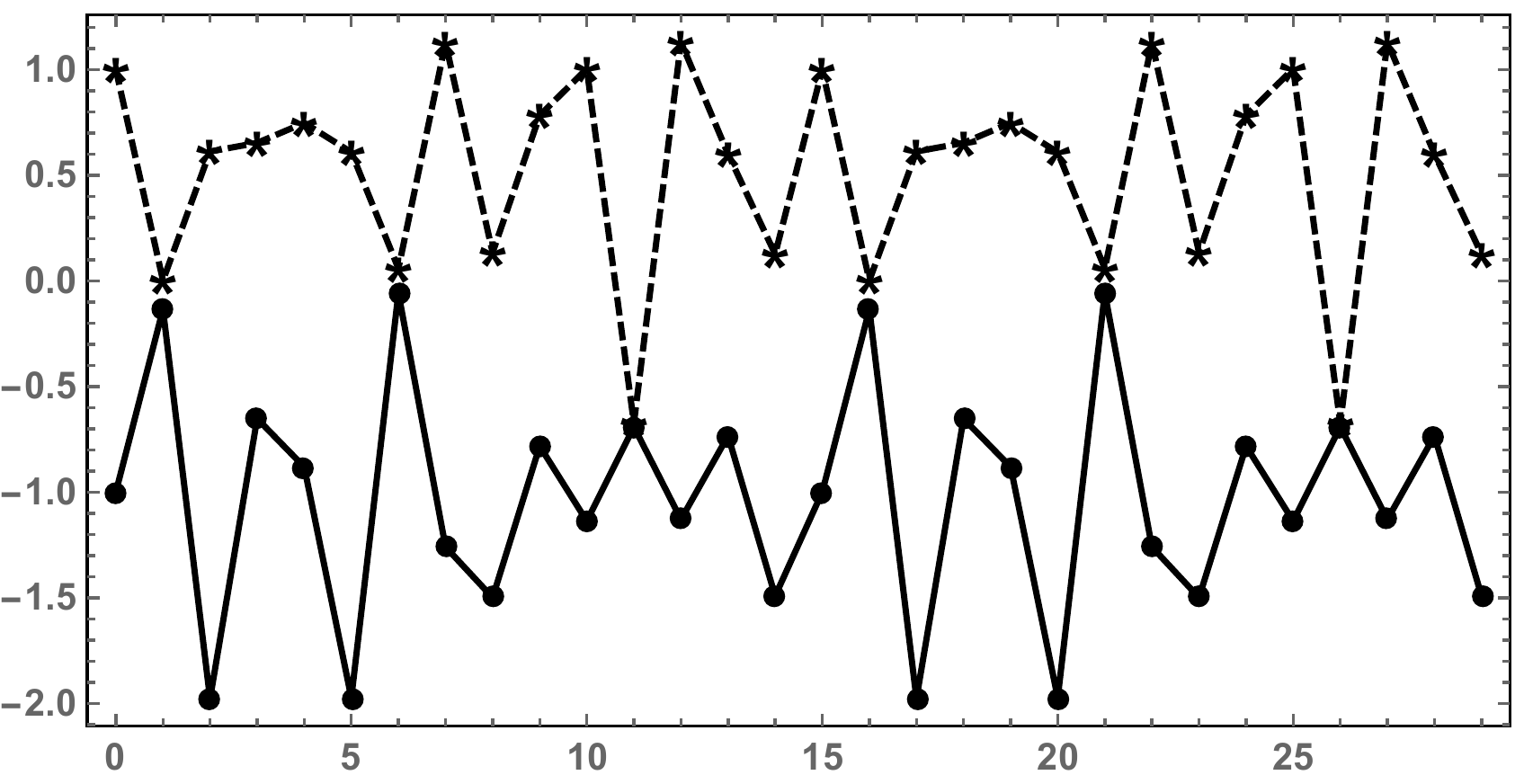} %
\includegraphics[width=8cm]{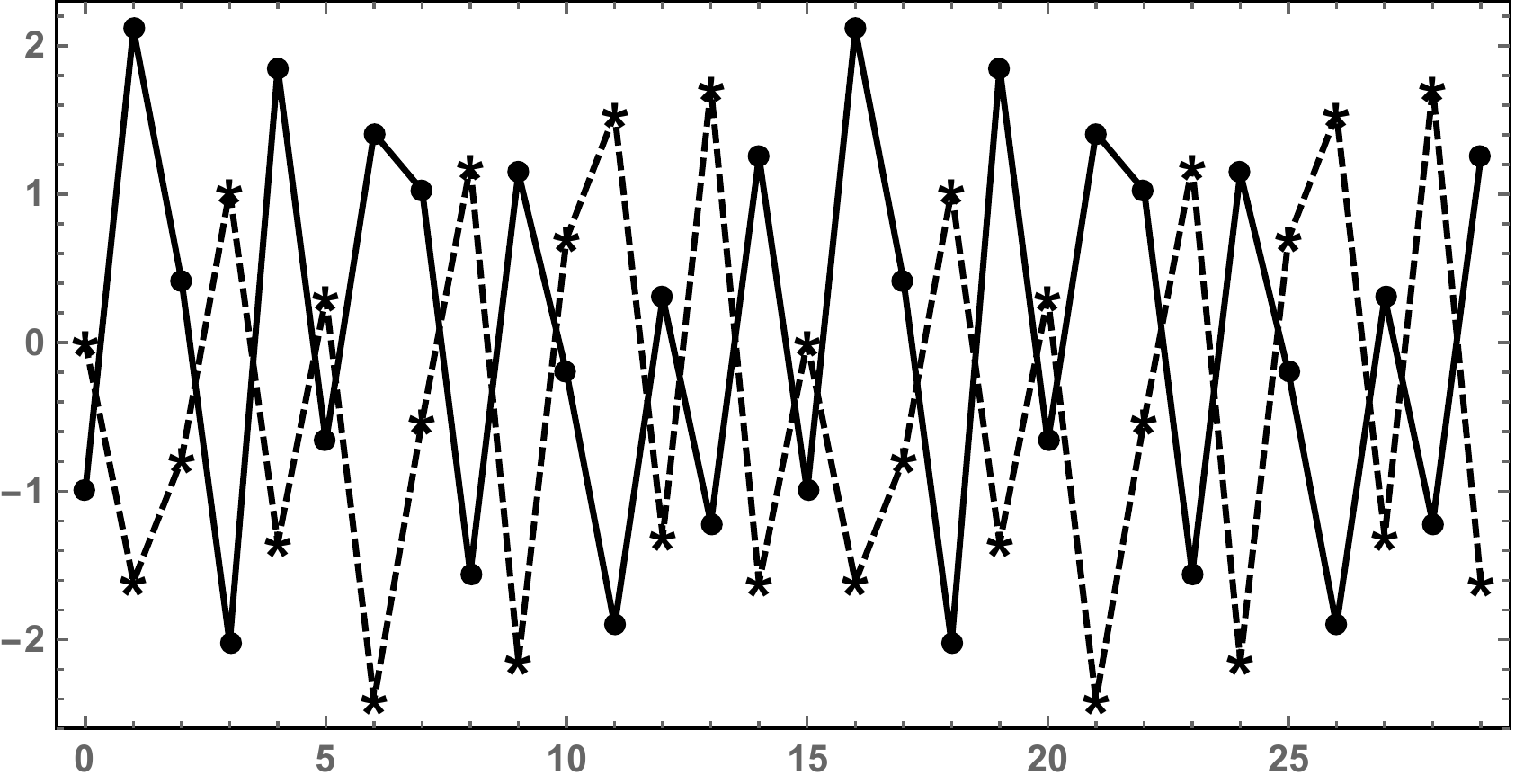}
\end{center}
\caption{\textbf{Example 1a.} Top graph: the evolution of the real parts $\mbox{Re}%
\left[x_{1}(\ell)\right]$ (dots) and $\mbox{Re}\left[x_{2}(\ell)\right]$
(stars). Bottom graph: the evolution of the imaginary parts $\mbox{Im}\left[%
x_{1}(\ell)\right]$ (dots) and $\mbox{Im}\left[x_{2}(\ell)\right]$ (stars).
The evolution is, of course, with respect to the discrete-time variable $%
\ell $, which corresponds to the horizontal axis. Note the periodicity with
 period $L=15$. }
\label{Fig:Ex1aReIm}
\end{figure}

\textbf{ Example 1a.} We begin with an \textit{isochronous} case of system~(\ref{SystGen0}) with the solution~(\ref{SolGen0}). In Figures~\ref{Fig:Ex1aBoth} and \ref{Fig:Ex1aReIm} we  plot the solution 
$\left\{ x_{1}(\ell ),x_{2}(\ell )\right\} $ of system (\ref{SystGen0}) with 
\begin{eqnarray}
&&a_{1}=\exp \left( \frac{2\pi \mathbf{i}}{3}\right) ~,~~~a_{2}=\exp \left( 
\frac{4\pi \mathbf{i}}{5}\right) ;~~b_{1}=1~,~~b_{2}=2~;  \notag \\
&&x_{1}(0)=-1-\mathbf{i}~,~~~x_{2}(0)=1~.  \label{Ex1aParameters}
\end{eqnarray}%
The solution of this system is \textit{periodic} with period $L=15$, the
Least Common Multiple of $3$ and $5$, see \textbf{Remark 3.1}.

\bigskip

\textbf{Example 1b.} This is an \textit{asymptotically isochronous} case
of system (\ref{SystGen0}) with the solution (\ref{SolGen0}). In Figure~\ref%
{Fig:Ex1bReIm} we  plot  the solution $\left\{ x_{1}(\ell
),x_{2}(\ell )\right\} $ of system~(\ref{SystGen0}) with 
\begin{eqnarray}
&&a_{1}=\exp \left( \frac{2\pi \mathbf{i}}{7}\right) ~,~~~a_{2}=0.9\exp
\left( \frac{4\pi \mathbf{i}}{5}\right) ~;~~b_{1}=.1~,~~b_2=.2~;  \notag \\
&&x_{1}(0)=-1-\mathbf{i}~,~~~x_{2}(0)=1~.  \label{Ex1bParameters}
\end{eqnarray}%
The solution of this system is \textit{asymptotically periodic} with
asymptotic period $L=7$. 

\begin{figure}[tbp]
\begin{center}
\includegraphics[width=8cm]{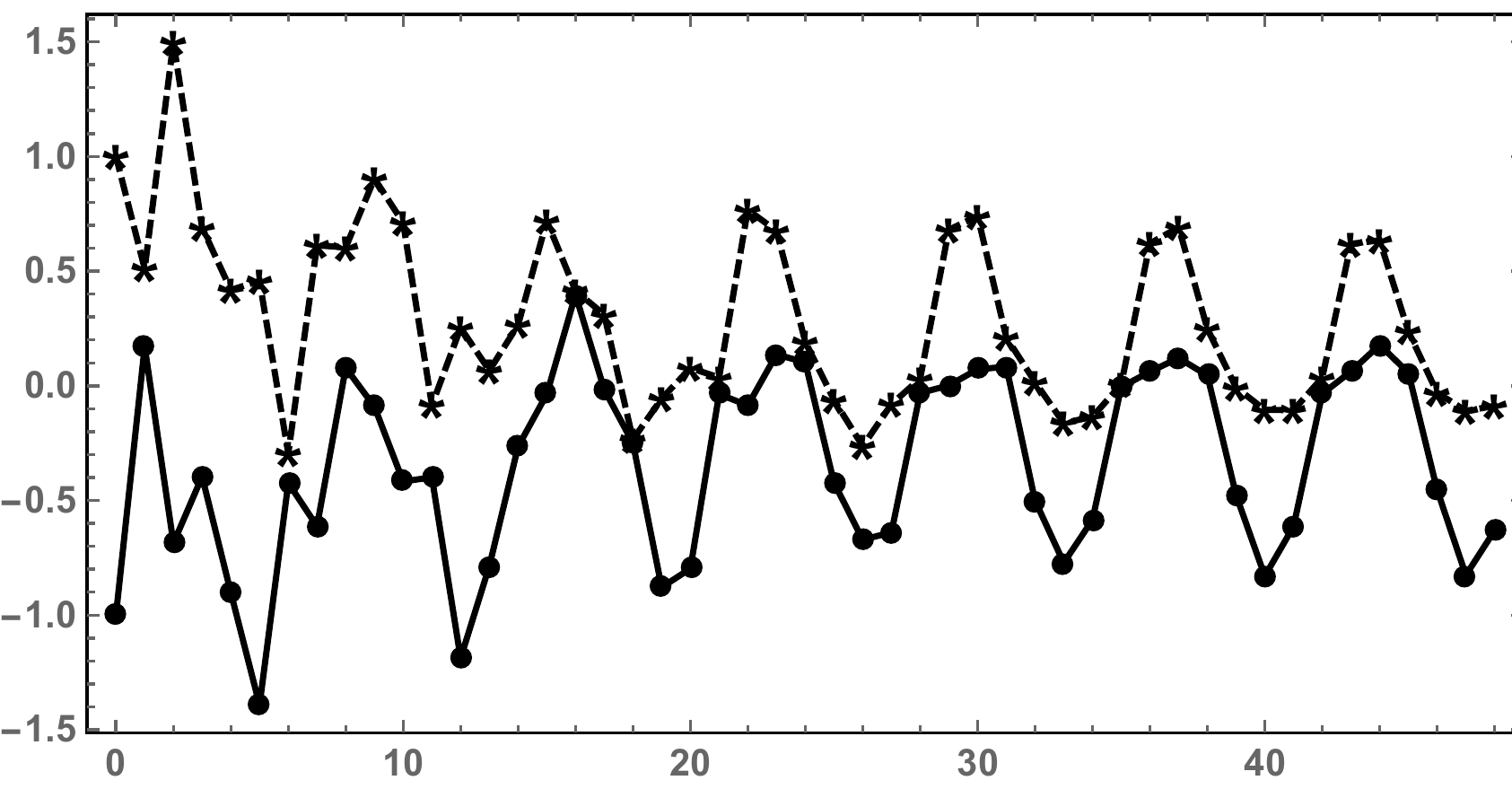} %
\includegraphics[width=8cm]{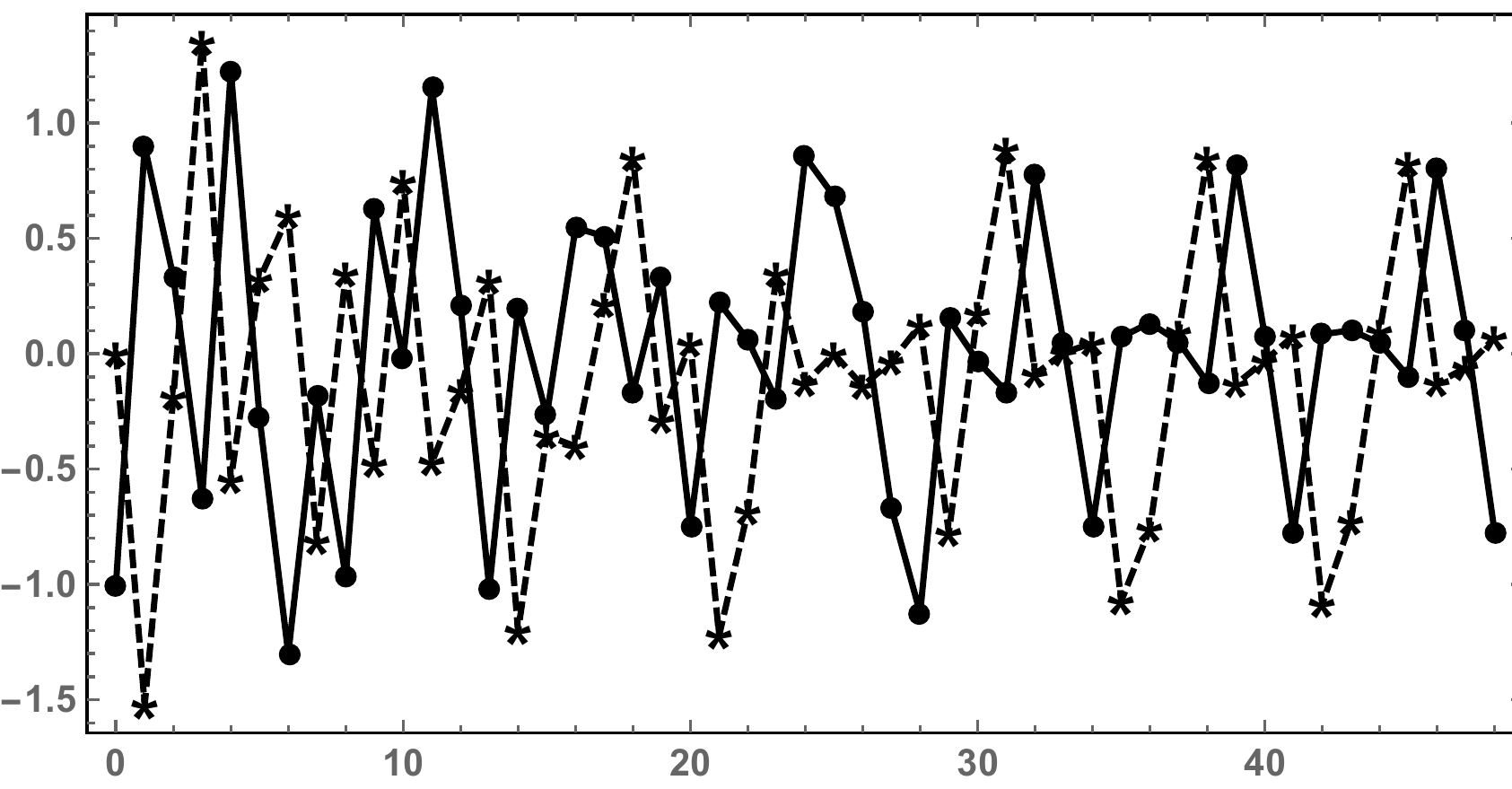}
\end{center}
\caption{\textbf{Example 1b}. Top graph: the evolution of the real parts $\mbox{Re}%
\left[x_{1}(\ell)\right]$ (dots) and $\mbox{Re}\left[x_{2}(\ell)\right]$
(stars). Bottom graph: the evolution of the imaginary parts $\mbox{Im}\left[%
x_{1}(\ell)\right]$ (dots) and $\mbox{Im}\left[x_{2}(\ell)\right]$ (stars).
The complex pairs~$x_{1}(\ell), x_2(\ell)$ are ordered \textit{lexicographically}. The evolution is, of course, with respect to
the discrete-time variable $\ell$, which corresponds to the horizontal axis.
Note the asymptotic periodicity with  asymptotic period $L=7$. Occasionally a dot and a start are too close for their separation to be actually visible.}
\label{Fig:Ex1bReIm}
\end{figure}

\begin{figure}[tbp]
\begin{center}
\includegraphics[width=10cm]{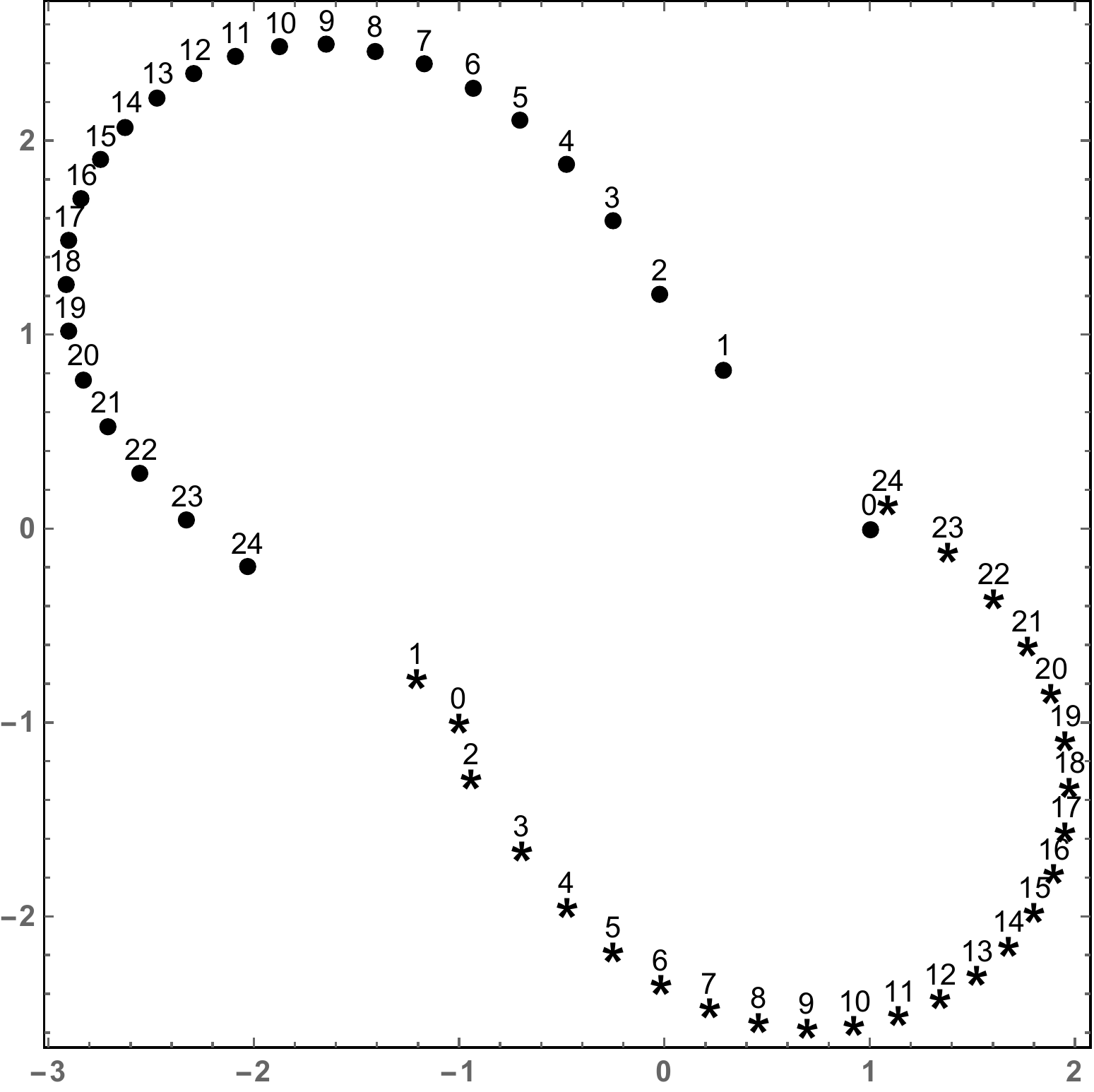}
\end{center}
\caption{\textbf{Example 1c}. For each $\ell=0,1,\ldots, 25$, the position of $%
x_{1}(\ell)$ in the \textit{complex} $x$-plane is indicated by a \textit{dot}
and the label $\ell$ that indicates the value of the discrete-time, while
the position of $x_{2}(\ell)$ is indicated by a \textit{star} and the
analogous label~$\ell$. The pairs~$(x_{1}(\ell),x_{2}(\ell))$ are ordered by
\textit{contiguity}.}
\label{Fig:Ex1cBoth}
\end{figure}

\begin{figure}[tbp]
\begin{center}
\includegraphics[width=8cm]{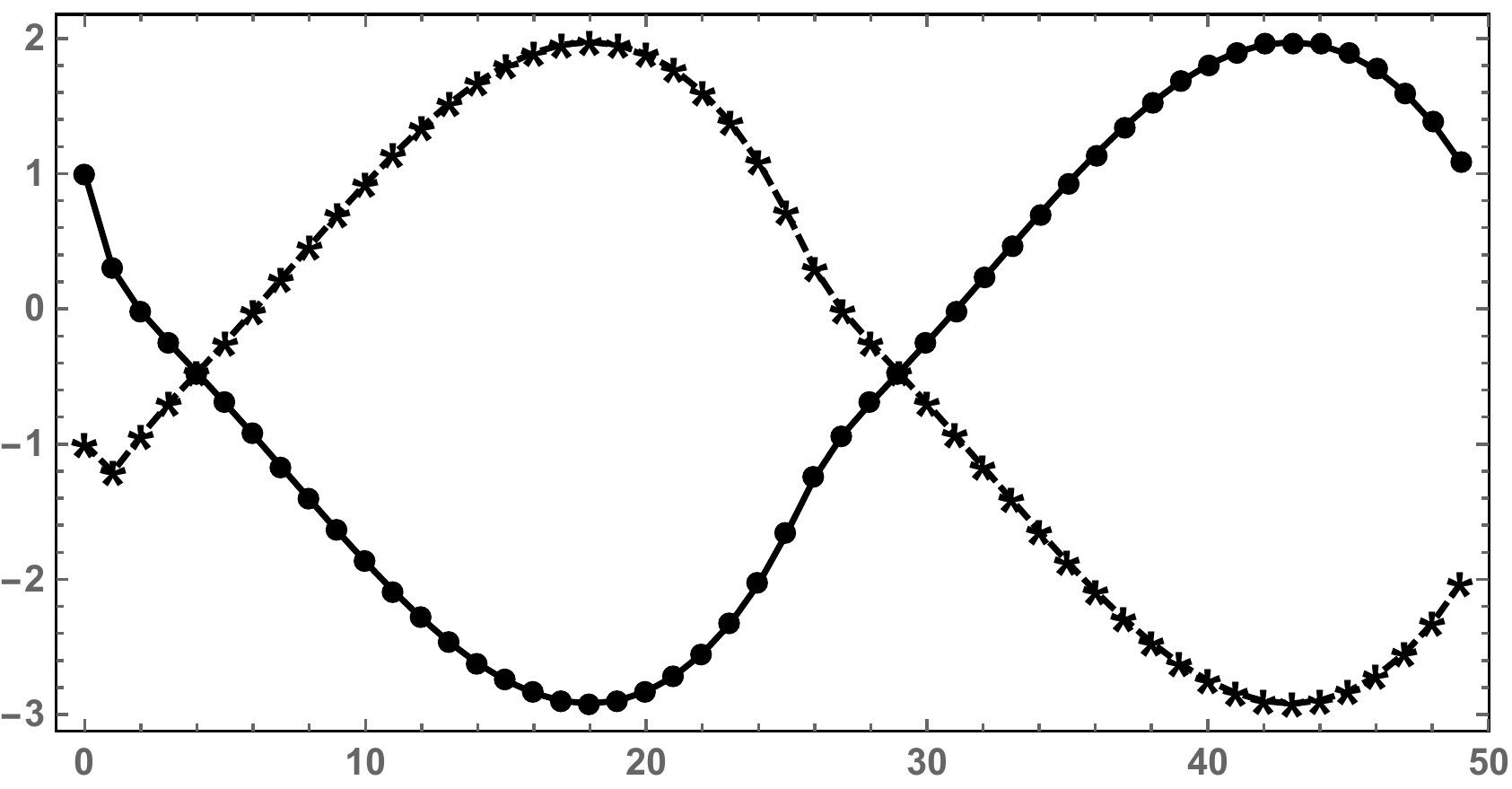} %
\includegraphics[width=8cm]{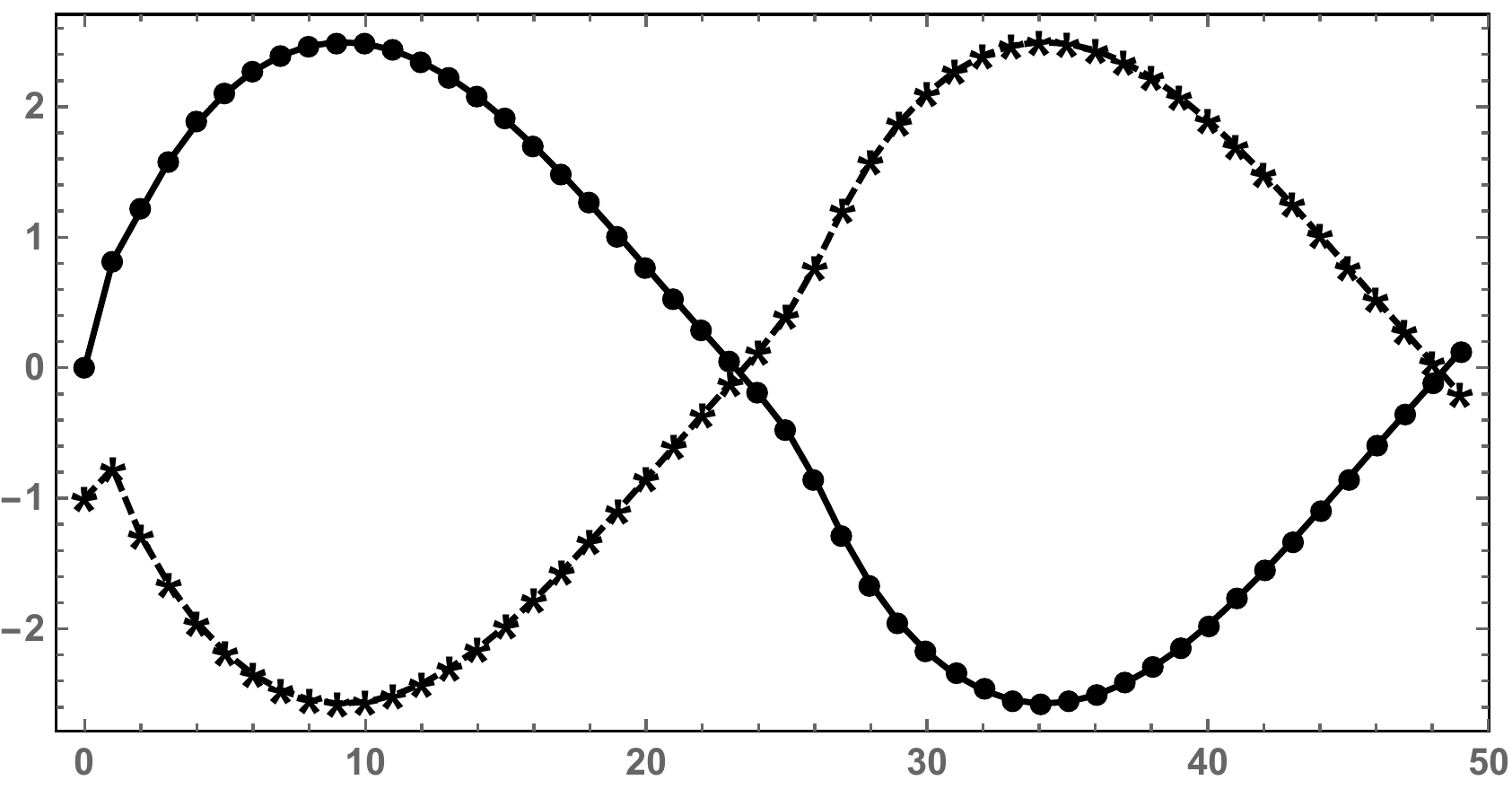}
\end{center}
\caption{\textbf{Example 1c}. Top graph: the evolution of the real parts $\mbox{Re}%
\left[x_{1}(\ell)\right]$ (dots) and $\mbox{Re}\left[x_{2}(\ell)\right]$
(stars). Bottom graph: the evolution of the imaginary parts $\mbox{Im}\left[%
x_{1}(\ell)\right]$ (dots) and $\mbox{Im}\left[x_{2}(\ell)\right]$ (stars).
The evolution is, of course, with respect to the discrete-time variable $%
\ell $, which corresponds to the horizontal axis. The pairs~$(x_{1}(\ell),x_{2}(\ell))$ are ordered by
\textit{contiguity}. Note the asymptotic
periodicity with period $L=25$. }
\label{Fig:Ex1cReIm}
\end{figure}

\textbf{ Example 1c.} We give another example of an \textit{asymptotically
isochronous} case of system~(\ref{SystGen0}) with the solution (\ref{SolGen0}%
). In this example we order the components of the solution \textit{by
contiguity}, see \textbf{Remark 2.3}. In Figures \ref{Fig:Ex1cBoth} and \ref{Fig:Ex1cReIm} we
plot the solution $\left\{ x_{1}(\ell ),x_{2}(\ell )\right\} 
$ of system (\ref{SystGen0}) with 
\begin{eqnarray}
&&a_{1}=0.1\exp \left( \frac{2\pi \mathbf{i}}{3}\right) ~,~~~a_{2}=\exp
\left( \frac{2\pi \mathbf{i}}{25}\right) ;~~b_{1}=1~,~~b_{2}=1~;  \notag \\
&&x_{1}(0)=-1-\mathbf{i}~,~~~x_{2}(0)=1~.  \label{Ex1aParameters}
\end{eqnarray}%
The solution of this system is asymptotically periodic with  asymptotic
period $L=25$.

\bigskip

\textbf{Example 2. \textit{Generation one} system with \textit{seed}~(\ref{SystSeed})}

\begin{figure}[tbp]
\begin{center}
\includegraphics[width=8cm]{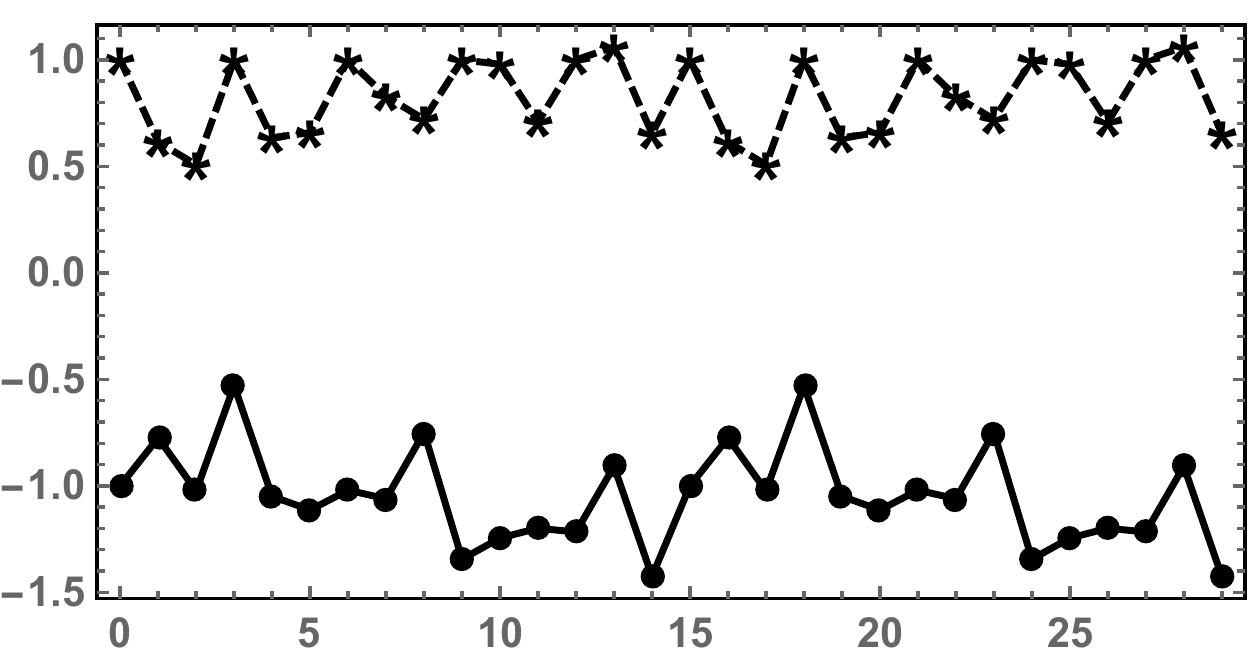} %
\includegraphics[width=8cm]{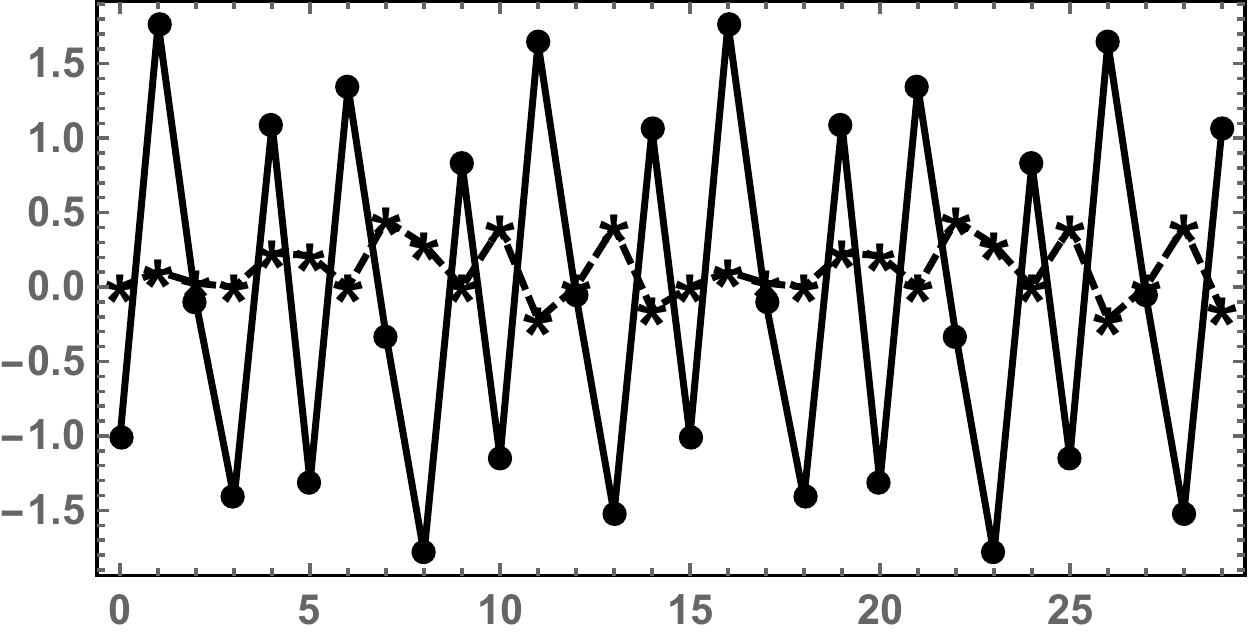}
\end{center}
\caption{\textbf{Example 2a}. Top graph: the evolution of the real parts $\mbox{Re}%
\left[x_{1}(\ell)\right]$ (dots) and $\mbox{Re}\left[x_{2}(\ell)\right]$
(stars). Bottom graph: the evolution of the imaginary parts $\mbox{Im}\left[%
x_{1}(\ell)\right]$ (dots) and $\mbox{Im}\left[x_{2}(\ell)\right]$ (stars).
The pairs~$(x_{1}(\ell),x_{2}(\ell))$ are ordered \textit{lexicographically}. The evolution is, of course, with respect to the discrete-time variable $%
\ell $, which corresponds to the horizontal axis. 
 Note the periodicity with
 period $L=15$. }
\label{Fig:Ex2aReIm}
\end{figure}

\textbf{ Example 2a.} We begin with an \textit{isochronous} case of 
\textit{generation~one} system~(\ref{SystGen1}) with the solution~(\ref{SolGen1}).
In Figure~\ref{Fig:Ex2aReIm} we  plot  the solution $\left\{
x_{1}(\ell ),x_{2}(\ell )\right\} $ of system~~(\ref{SystGen1})  with 
\begin{eqnarray}
&&a_{1}=\exp \left( \frac{2\pi \mathbf{i}}{3}\right) ~,~~~a_{2}=\exp \left( 
\frac{4\pi \mathbf{i}}{5}\right) ;~~b_{1}=1~,~~b_{2}=2~;  \notag \\
&&x_{1}(0)=-1-\mathbf{i}~,~~~x_{2}(0)=1~.  \label{Ex2aParameters}
\end{eqnarray}%
The solution of this system is periodic with period $L=15$, which is the
Least Common Multiple of $3$ and $5$, see \textbf{Remark 3.1}.

\bigskip

\begin{figure}[tbp]
\begin{center}
\includegraphics[width=8cm]{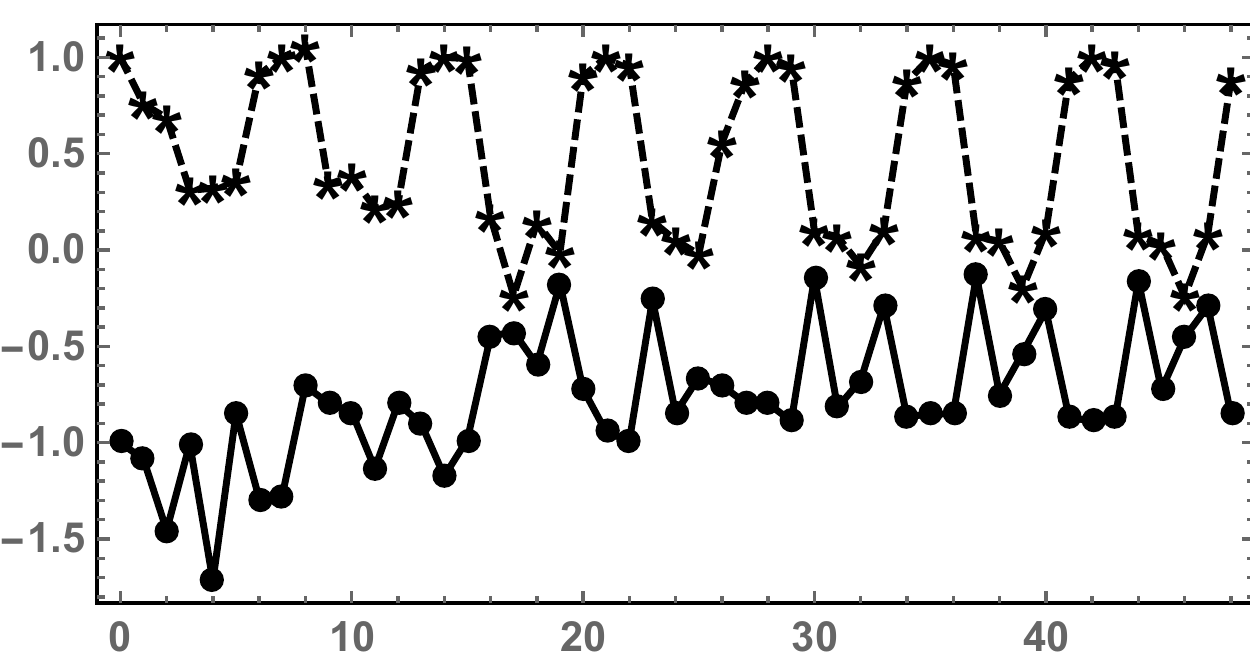} %
\includegraphics[width=8cm]{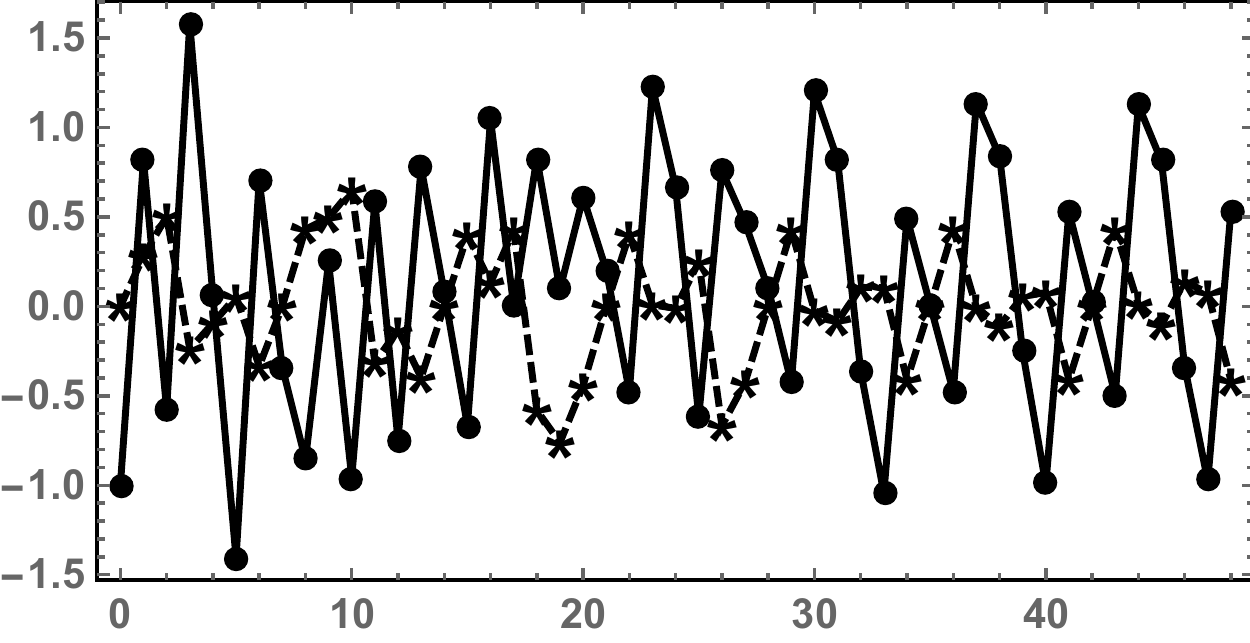}
\end{center}
\caption{\textbf{Example 2b}. Top graph: the evolution of the real parts $\mbox{Re}%
\left[x_{1}(\ell)\right]$ (dots) and $\mbox{Re}\left[x_{2}(\ell)\right]$
(stars). Bottom graph: the evolution of the imaginary parts $\mbox{Im}\left[%
x_{1}(\ell)\right]$ (dots) and $\mbox{Im}\left[x_{2}(\ell)\right]$ (stars).
The pairs~$(x_{1}(\ell),x_{2}(\ell))$ are ordered \textit{lexicographically}.
The evolution is, of course, with respect to the discrete-time variable $%
\ell $, which corresponds to the horizontal axis. 
Note the asymptotic
periodicity with  asymptotic period $L=7$. }
\label{Fig:Ex2bReIm}
\end{figure}

\textbf{ Example 2b.} This is an \textit{asymptotically isochronous} case
of generation~one system~(\ref{SystGen1}) with the solution~(\ref%
{SolGen1}). In Figure~\ref{Fig:Ex2bReIm} we  plot the solution $%
\left\{x_{1}(\ell ),x_{2}(\ell )\right\}$ of system~(\ref{SystGen1}) with 
\begin{eqnarray}
&&a_{1}=\exp \left( \frac{2\pi \mathbf{i}}{7}\right) ~,~~~a_{2}=0.9 \exp
\left( \frac{4\pi \mathbf{i}}{5}\right)~; ~~b_{1}=.1~,~~b_2=.2~;  \notag \\
&&x_{1}(0)=-1-\mathbf{i}~,~~~x_{2}(0)=1~.  \label{Ex2bParameters}
\end{eqnarray}%
The solution of this system is asymptotically periodic with  asymptotic
period $L=7$.

\bigskip

\textbf{ Example 3. \textit{Generation two} system with \textit{seed}~(\ref{SystSeed})}

\begin{figure}[tbp]
\begin{center}
\includegraphics[width=8cm]{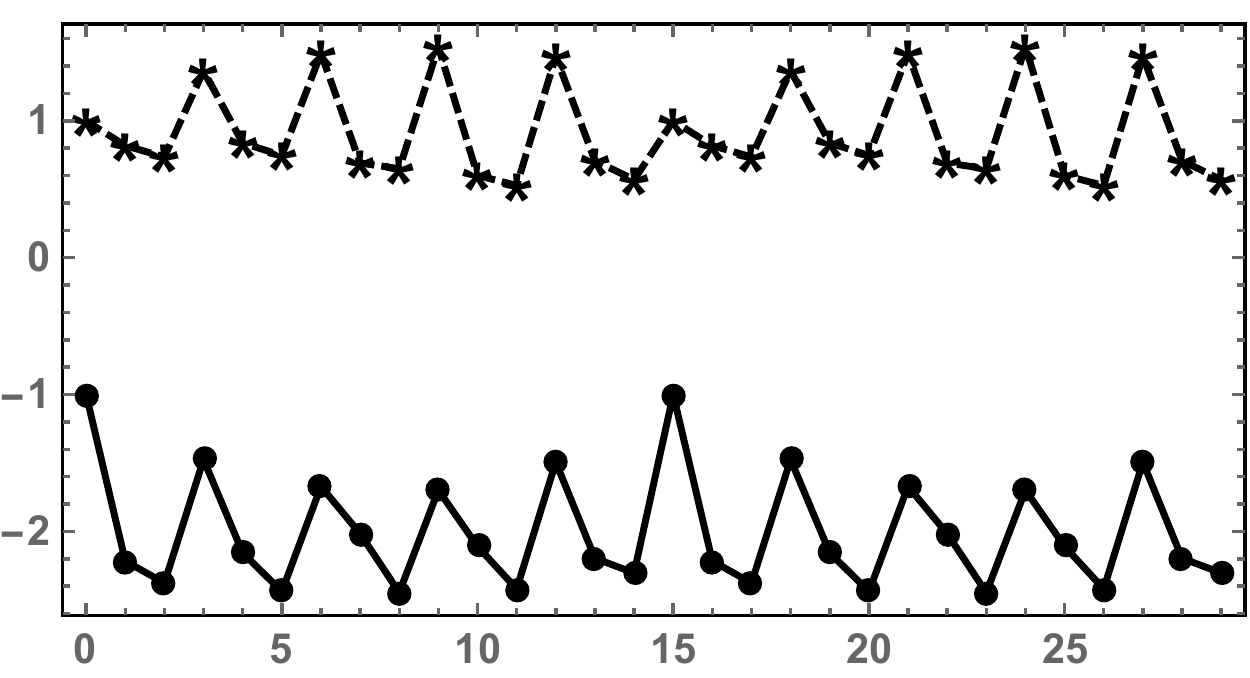} %
\includegraphics[width=8cm]{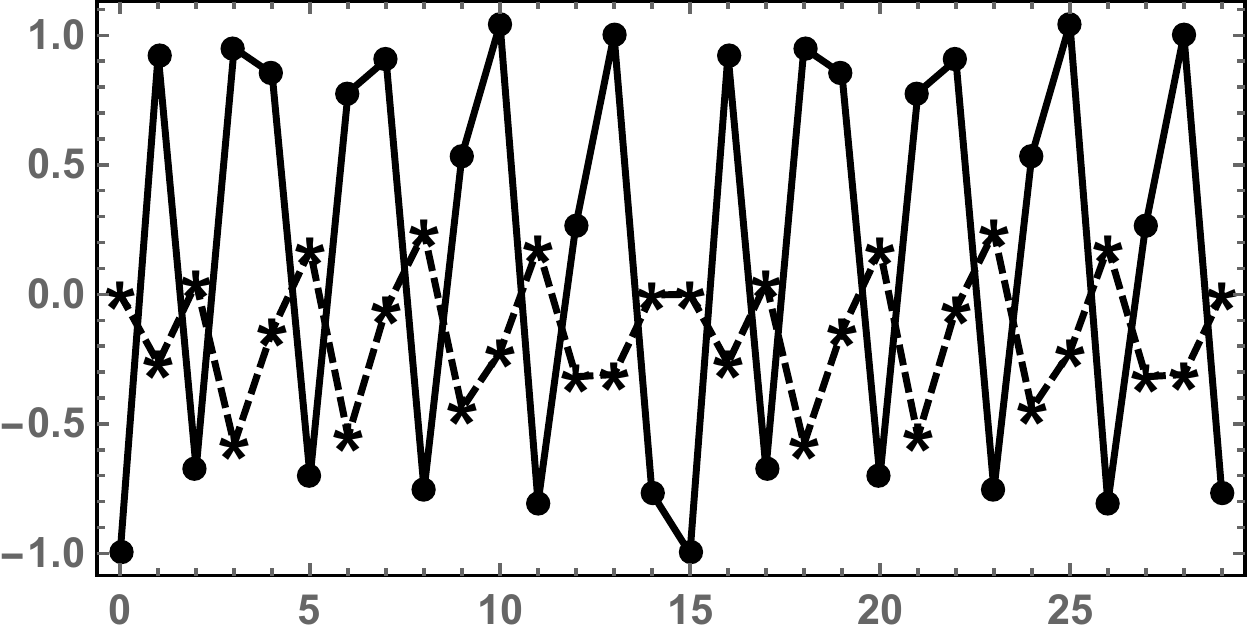}
\end{center}
\caption{ \textbf{Example 3a}. Top graph: the evolution of the real parts $\mbox{Re}%
\left[x_{1}(\ell)\right]$ (dots) and $\mbox{Re}\left[x_{2}(\ell)\right]$
(stars). Bottom graph: the evolution of the imaginary parts $\mbox{Im}\left[%
x_{1}(\ell)\right]$ (dots) and $\mbox{Im}\left[x_{2}(\ell)\right]$ (stars).
The evolution is, of course, with respect to the discrete-time variable $%
\ell $, which corresponds to the horizontal axis. Note the periodicity with
period $L=15$. }
\label{Fig:Ex3aReIm}
\end{figure}

\textbf{ Example 3a.} We begin with an \textit{isochronous} case of the
\textit{generation~two} system derived from the \textit{seed system}~(\ref{SystSeed})
according to the general procedure described in Section~\ref{sec2} and
discussed in Section~\ref{sec3} for the case of the  \textit{seed system}~(\ref{SystSeed}) (with $N=2$). 
In Figure~\ref{Fig:Ex3aReIm} we  plot the
solution $\left\{ x_{1}(\ell ),x_{2}(\ell )\right\} $ of the generation two
system with 
\begin{eqnarray}
&&a_{1}=\exp \left( \frac{2\pi \mathbf{i}}{3}\right) ~,~~~a_{2}=\exp \left( 
\frac{4\pi \mathbf{i}}{5}\right) ;~~b_{1}=1~,~~b_{2}=2~;  \notag \\
&&x_{1}(0)=-1-\mathbf{i}~,~~~x_{2}(0)=1~.  \label{Ex3aParameters}
\end{eqnarray}%
The solution of this system is periodic with  period $L=15$, which is the
Least Common Multiple of $3$ and $5$, see \textbf{Remark 3.1}.

\bigskip

\begin{figure}[tbp]
\begin{center}
\includegraphics[width=8cm]{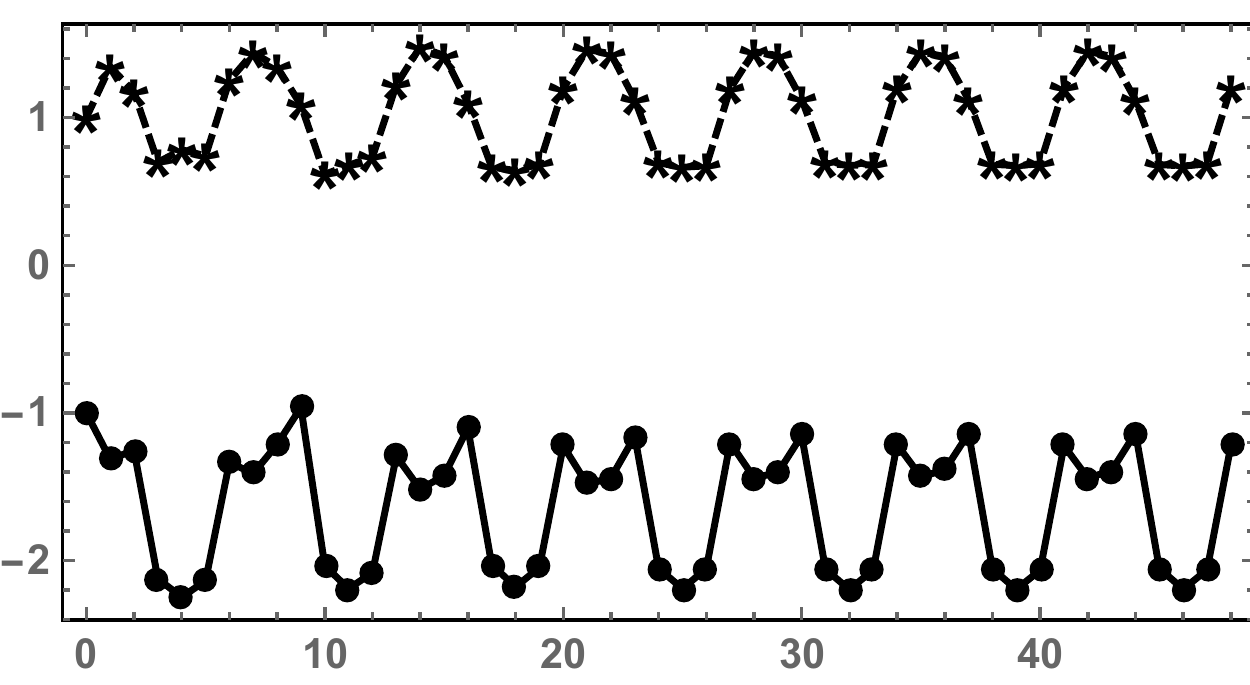} %
\includegraphics[width=8cm]{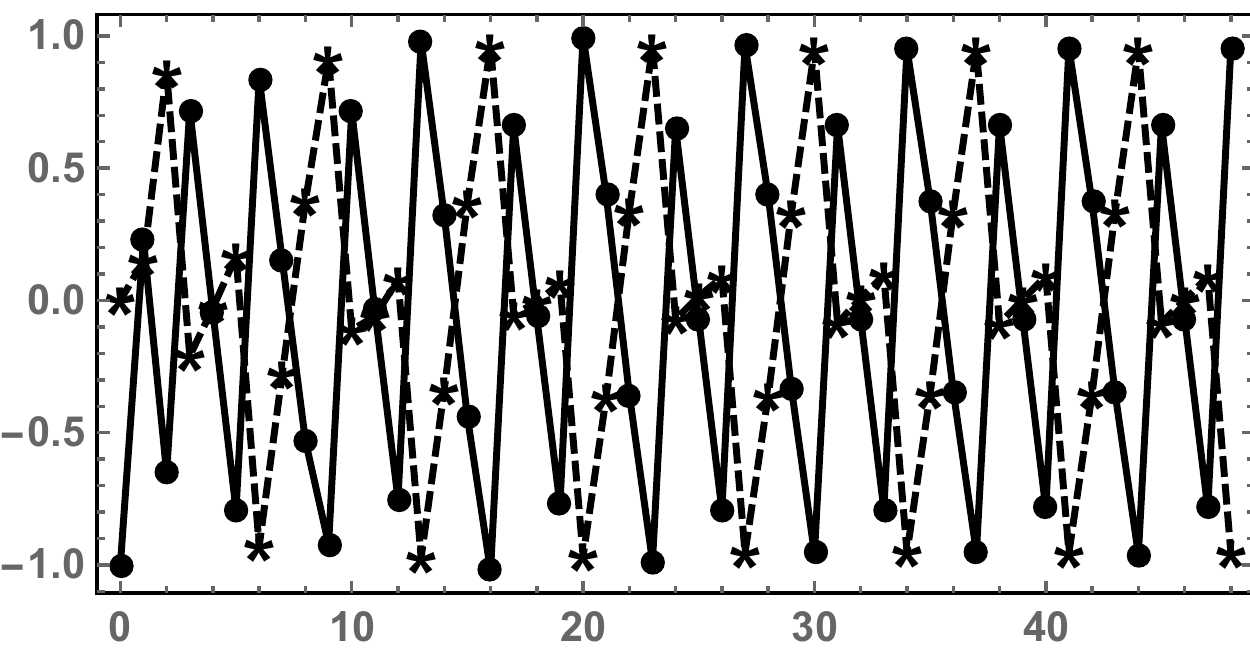}
\end{center}
\caption{\textbf{Example 3b}. Top graph: the evolution of the real parts $\mbox{Re}%
\left[x_{1}(\ell)\right]$ (dots) and $\mbox{Re}\left[x_{2}(\ell)\right]$
(stars). Bottom graph: the evolution of the imaginary parts $\mbox{Im}\left[%
x_{1}(\ell)\right]$ (dots) and $\mbox{Im}\left[x_{2}(\ell)\right]$ (stars).
The evolution is, of course, with respect to the discrete-time variable $%
\ell $, which corresponds to the horizontal axis. The pair $(x_1(\ell), x_2(\ell))$ is ordered
\textit{lexicographically}. Note the asymptotic
periodicity with  asymptotic period $L=7$. }
\label{Fig:Ex3bReIm}
\end{figure}

\textbf{ Example 3b.} This is an \textit{asymptotically isochronous} case
of the \textit{generation~two} system derived from the \textit{seed system} (\ref%
{SystSeed}) according to the general procedure described in Section 2 and
discussed in Section 3 for the case of the  \textit{seed system}~(\ref{SystSeed}) 
(with $N=2$). In Figure \ref{Fig:Ex3bReIm} we plot  the
solution $\left\{ x_{1}(\ell ),x_{2}(\ell )\right\} $ of the \textit{generation two} system  with 
\begin{eqnarray}
&&a_{1}=\exp \left( \frac{2\pi \mathbf{i}}{7}\right) ~,~~~a_{2}=0.9\exp
\left( \frac{4\pi \mathbf{i}}{5}\right) ~;~~b_{1}=.1~,~~b_2=.2~;  \notag \\
&&x_{1}(0)=-1-\mathbf{i}~,~~~x_{2}(0)=1~.  \label{Ex3bParameters}
\end{eqnarray}%
The solution of this system is \textit{asymptotically periodic} with
asymptotic period $L=7$.

\begin{figure}[tbp]
\begin{center}
\includegraphics[width=8cm]{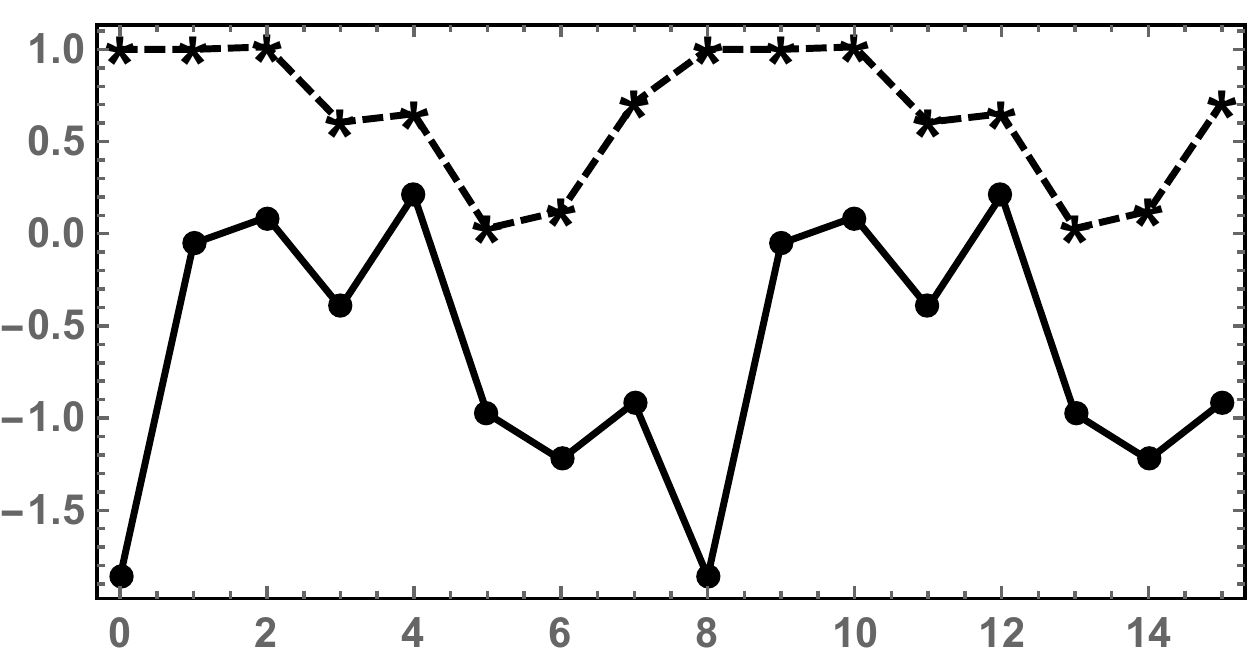} \includegraphics[width=8cm]{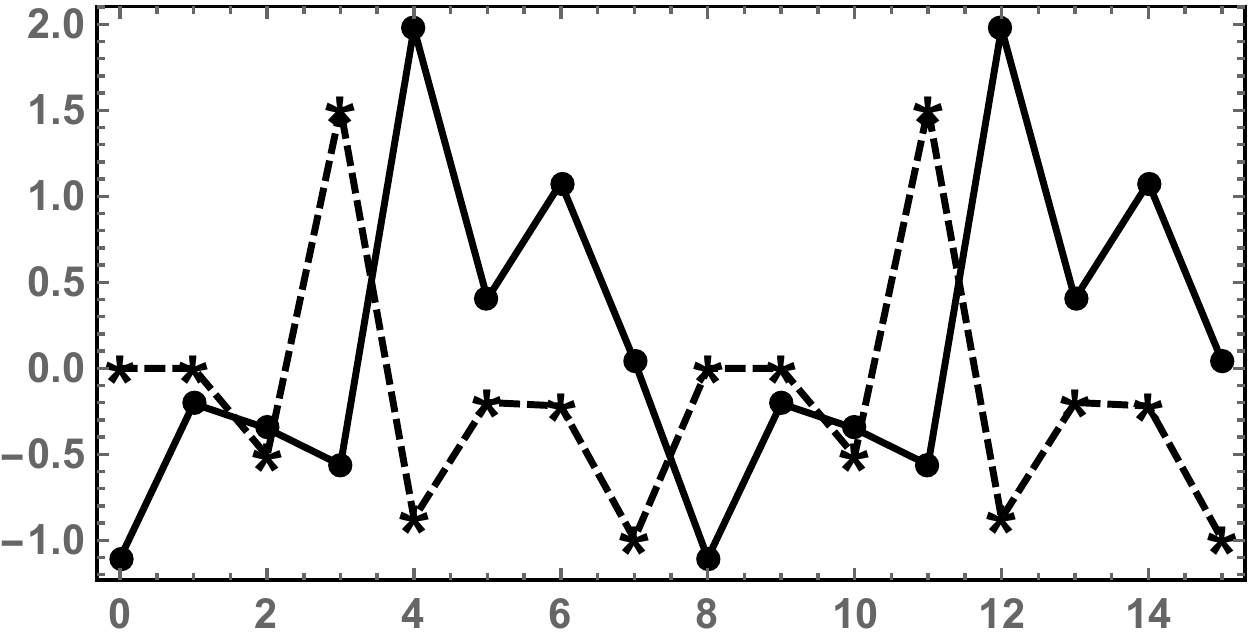}
\end{center}
\caption{\textbf{Example 4.} Top graph: the evolution of the real parts $\mbox{Re}%
\left[x_{1}(\ell)\right]$ (dots) and $\mbox{Re}\left[x_{2}(\ell)\right]$
(stars). Bottom graph: the evolution of the imaginary parts $\mbox{Im}\left[%
x_{1}(\ell)\right]$ (dots) and $\mbox{Im}\left[x_{2}(\ell)\right]$ (stars).
The pair $(x_1(\ell), x_2(\ell))$ is ordered
\textit{lexicographically}. The evolution is, of course, with respect to
the discrete-time variable $\ell$, which corresponds to the horizontal axis.
Note the periodicity with period $L=8$. }
\label{Fig:Ex4ReIm}
\end{figure}

\textbf{Example 4}. In Figure~\ref{Fig:Ex4ReIm} we plot the solutions of
system~(\ref{SystEx4x}) with (\ref{ambmaut}), (\ref{amPeriodic}), for the case where $N=2$ and 
\begin{equation}
a_{1}=\exp ({\mathbf{i~}}\pi )~,~~~a_{2}=\exp \left( \frac{{\mathbf{i~}}\pi 
}{2}\right) ~;~~~b_{1}=1,~~~b_{2}=2~.  \label{Ex4Parameters}
\end{equation}%
We used conditions~(\ref{PeriodicityConditions}) with $\beta _{1}=\exp ({%
\mathbf{i~}}\pi )$ and $\beta _{2}=\exp ({\mathbf{i~}}\pi )$ to find
initial conditions for which system~(\ref{SystEx4x}) with (\ref{Ex4Parameters})
has a periodic solution:%
\begin{eqnarray}
&&x_{1}(0)=-\frac{17^{1/4}+\gamma }{17^{1/4}+(1+2~\mathbf{i})~\gamma
-2~(-3)^{1/4}~\gamma },  \notag \\
&&x_{1}(1)=\left[ 1+\mathbf{i}-3^{1/4}~\exp \left( \frac{\mathbf{i~}\pi }{4}%
\right) \right] ~x_{1}(0),  \notag \\
&&x_{2}(0)=1~,~~~x_{2}(1)=1~,  \label{Ex4InitCond}
\end{eqnarray}%
where $\gamma =\exp \left[ \mathbf{i}\arctan (4)/2)\right] $. The solution
of system~(\ref{SystEx4x}) with (\ref{Ex4Parameters}) with the initial
conditions~(\ref{Ex4InitCond}) is periodic with period $8$.

\end{document}